\documentclass[fleqn]{iopart}

\usepackage{times}
\usepackage{epsfig}
\usepackage{graphicx,iopams,setstack,color}

\font\bfgreek=cmmib10
\def\bbu{{\hbox{\bfgreek\char'165}}}
\def\bbw{{\hbox{\bfgreek\char'167}}}
\def\bbs{{\hbox{\bfgreek\char'163}}}
\def\bbf{{\hbox{\bfgreek\char'146}}}

\def\rmd{{\rm d}}

\def\bea{\begin{eqnarray}}
\def\eea{\end{eqnarray}}

\def\had{{\sc had }}

\begin{document}

\title{Relativistic MHD with Adaptive Mesh Refinement}

\author{Matthew Anderson$^{1}$, 
Eric W.{} Hirschmann$^{2}$, Steven L.{} Liebling$^{3}$, David Neilsen$^{2}$}

\address{
$1$ Department of Physics and Astronomy, Louisiana State
  University, Baton Rouge, LA 70803-4001, USA\\
$2$ Department of Physics and Astronomy, Brigham Young University,
  Provo, UT 84602, USA\\
$3$ Department of Physics, Long Island University -- C.W. Post Campus,
  Brookville, NY 11548, USA
}

%
%

\begin{abstract}
This paper presents a new computer code to solve the general relativistic
magnetohydrodynamics~(GRMHD) equations using distributed parallel adaptive 
mesh refinement (AMR).  The fluid equations are solved using a 
finite difference Convex ENO method~(CENO) in $3+1$ dimensions, and the
AMR is Berger-Oliger.  Hyperbolic divergence cleaning is used to control
the $\nabla\cdot {\bf B}=0$ constraint.  We present results from three flat 
space tests, and examine the accretion of a fluid onto a Schwarzschild black 
hole, reproducing the Michel solution.
The AMR simulations substantially improve
performance while reproducing the resolution equivalent unigrid simulation 
results.  Finally, we discuss strong scaling results for parallel unigrid
and AMR runs.
\end{abstract}


%
%
\section{Introduction}
\label{sec:introduction}

The interaction of gravitational and electromagnetic fields together with
rotation is believed to power the central engines of many astrophysical
phenomena including relativistic jets in 
active galactic nuclei (AGN), other forms of black
hole accretion, gamma-ray bursts (GRB), and core collapse supernovae.
In addition, interactions between strong gravitational and
electromagnetic fields are believed to result
in the transport of angular momentum in accretion disks
via the magnetorotational instability (MRI)
and in the extraction of black hole energy via the Blandford-Znajek mechanism.

The expected ubiquity of magnetic fields in the vicinity of strongly
gravitating
compact objects has spurred increased theoretical efforts to understand
these astrophysical systems.  The difficulties, however, are formidable as the
physical laws describing these phenomena are nonlinear, evolutionary, and, in
general, without simplifying symmetries.  As a consequence, numerical simulation
of these systems becomes crucial for better understanding them.  Even then,
there are likely considerable aspects of the physics that remain out of reach
of current computational resources.  For instance, it is 
not unrealistic to imagine that
dissipative and radiative aspects of these problems will be important for
accurately modeling certain types of phenomena.  However, with the hope of
capturing some of the relevant physics, a number of groups have begun
developing methods and codes for
evolving some of the relativistic components of these complicated scenarios.

In this large and growing body of work, a significant effort is directed at
special relativistic magnetohydrodynamics (RMHD).  However, with
increasing interest in the interaction of gravitational and
electromagnetic fields, one must also
couple the equations of relativistic MHD to general relativity either through
a curved space background or the dynamical field equations themselves.
The earliest attack on this general problem was Wilson's pioneering work
evolving a rotating, axisymmetric star with a poloidal magnetic
field~\cite{Wilson1975}.  Subsequently, little numerical work was done until very recently with
several groups developing codes for evolving the general relativistic
MHD (GRMHD) equations on fixed
backgrounds ~\cite{Koide:2000, DeVilliers:2002ab, Gammie:2003rj, Baumgarte:2002b, Komissarov:2004, Anton:2005gi, Anninos:2005} and in dynamical spacetimes
~\cite{Duez:2005sf,Shibata:2005gp}.

One of the difficulties alluded to above in the numerical simulation of
these sorts of systems is that one must solve the GRMHD equations
over a large range of time and length scales.  The computational
requirements necessary to adequately resolve multiscale phenomena
using only a single resolution mesh are often too high for available resources.
This is particularly true when the full, dynamical GRMHD equations are being
considered.  Adaptive mesh refinement (AMR) therefore becomes crucial in
order to reduce the computational requirements necessary for modeling
such systems.

There are currently MHD codes in use which incorporate AMR, but most are
focused on the nonrelativistic 
problem~\cite{Balsara2001,Gombosi2000,Gombosi2001}.
Notable among these are the work of Balsara which extended
a version of constrained transport to AMR.
For relativistic MHD, only the work by Anninos~\etal~\cite{Anninos:2005} 
incorporates adaptive mesh refinement.  Theirs is a finite volume approach
with divergence cleaning.

This paper describes our algorithm for solving the GRMHD
equations with AMR.  Building on the work presented in~\cite{Neilsen2005},
where the RMHD equations were solved on overlapping grids,
some key elements of our algorithm are: (1) The
Convex Essentially Non-Oscillatory (CENO) method for the MHD equations,
(2) Hyperbolic divergence cleaning for controlling the
solenoidal constraint, (3) Berger-Oliger AMR, (4) Weighted Essentially
Non-Oscillatory (WENO) interpolation
for communications from coarse to fine grids, and (5) discretization in 
time via method of lines.

The CENO scheme is robust, and has three advantages for our code.  First,
the method does not require the spectral decomposition of the Jacobian
matrix, making it relatively efficient.  Central and central-upwind
schemes are
known to give results nearly identical to those of more complicated methods
for many problems~\cite{Lucas-Serrano:2004aq,Shibata:2005jv}. 
With the added capability of AMR, 
we find that we are able to efficiently resolve very fine solution features.
A second advantage of CENO for our purposes is that it is a finite difference,
or vertex centered, method.  As we will solve the Einstein equations with 
finite differences, using a finite difference fluid scheme simplifies coupling
the two sets of equations with AMR.  The simplification arises because
fluid and geometric variables are always defined at the same point as
grids are refined.  Finite volume and finite difference grids become
staggered with respect to each other  as they are refined. 
While we find this simplification 
advantageous in our present work, we note that our AMR code {\sc had}
can combine finite difference and finite volume schemes~\cite{motl2006}.
A third advantage is that ENO schemes are easily extended to
higher order accuracy.  Our CENO scheme can reconstruct fluid variables
to both first and second order, resulting in second and third order 
evolution schemes, respectively.

We choose hyperbolic divergence cleaning~\cite{Dedner2002}  to limit growth in
the solenoidal constraint on the magnetic field, $\nabla\cdot {\bf B}=0$.
Hyperbolic divergence cleaning is simple to implement.  A single
hyperbolic field is added to the system and coupled 
to the evolution equations for $\bf B$. Divergence cleaning gave
good results in earlier 
tests~\cite{Neilsen2005} and allows us to freely choose prolongation
methods for AMR.  While $\nabla\cdot {\bf B}=0$ is not satisfied to
machine precision for any particular discrete divergence operator, we find
that $||\nabla\cdot {\bf B}||$ does converge to zero, and, in the tests
presented here, $||\nabla\cdot {\bf B}||$ is roughly the same order
of magnitude as the expected truncation error.  

Mesh refinement is necessary for obtaining accurate numerical
solutions for three dimensional systems in  general relativity, and
AMR is essential for complicated problems where the refinement
regions can not be guessed {\it a priori.} We use the \had
infrastructure to provide Berger-Oliger style AMR~\cite{Berger}.  \had
has a modular design allowing one to easily implement many
different sets of evolution equations, and different modules can
be combined, for example, to solve both the Einstein and MHD equation
simultaneously.  \had supports higher order differencing
schemes, and our implementation of the MHD equations is fully third
order accurate~\cite{Lehner2006}.  Refinement regions can be specified in a
variety of ways. \had provides a shadow hierarchy for specifying
refinement criteria using truncation error estimates, or the user
may specify problem specific criteria, such as refining on gradients
or other solution features.  \had supports different
interpolation schemes (we choose WENO interpolation for this
work) and supports both finite volume and finite difference equations
or combinations of both.  Finally, as discussed below, \had
scales well in strong scaling tests in both unigrid and 
AMR tests.

Finally, we discretize the continuum equations first in space 
(creating a semi-discrete system) and then discretize in time using
the method of lines.  This gives us considerable flexibility in choosing
discretization schemes appropriate for very different types of 
equations.
For example, the MHD equations are solved here with
high-resolution shock-capturing methods, while we might solve the
Einstein equations using methods that preserve a discrete energy norm.
Using the method of lines, we can easily and consistently combine these
two sets of semi-discrete equations in a uniform time integration.
Time integrators can be independently chosen for their desired properties 
or order of integration.  For example, we choose for this work a
third-order Runge--Kutta scheme that preserves the TVD 
condition~\cite{ShuOsherI}.

The remaining sections of this paper give further details to the
algorithm sketched above and present code tests in both flat space
and on a Schwarzschild black hole background.  We first present the
GRMHD equations used in this work.

%
%
\section{The MHD equations in general relativity}
\label{subsec:equations}

A number of derivations of the GRMHD equations have appeared in the literature, 
e.g.,~\cite{Baumgarte:2002a,Sloan:1985,Evans:1988,Zhang:1989,Gammie:2003rj,
DeVilliers:2002ab,Neilsen2005}, and thus we simply present the equations to be
solved here.  The numerical methods that we use are very similar to those 
we have used in our previous MHD work using overlapping grids.  Thus our 
presentation here is short, and the reader may refer to that work for more 
detail~\cite{Neilsen2005}.

The spacetime metric is written in terms of the conventional ADM 3+1 
variables, namely 
\begin{equation}
\rmd s^2 = -\alpha^2 \,\rmd t^2 
+ h_{ij}(\rmd x^i + \beta^i \,\rmd t)(\rmd x^j + \beta^j \,\rmd t),
\end{equation}
where $\alpha$ is the lapse, $\beta^i$ is the shift, and $h_{ij}$ is the
$3$-metric on the spacelike hypersurfaces.  Units are chosen such that $c=1$ 
and $G=1$.  We denote the extrinsic curvature as $K_{ab}$ and the Christoffel 
coefficients with respect to the $3$-metric as ${^{3}{\Gamma}}_{ab}^i$.  As 
our focus in this paper is fixed background geometries, we will omit here a 
discussion of our approach to evolving the Einstein equations and only present
the relevant matter equations.  Future papers will address dynamical 
spacetimes.  

The equations for MHD on a curved background, as in flat spacetime, can 
be written, for the most part, in balance law form, namely
\begin{equation}
\partial_t\bbu  + \partial_k\bbf\,^k(\bbu) = \bbs(\bbu).
\label{eq:balance}
\end{equation}
where $\bbu$ is a state vector, $\bbf\,^k$ are flux functions, and $\bbs$ 
are source terms.  For the current case, these are 
\begin{eqnarray}
\fl \partial_t \left( \sqrt{h} \, D \right) + \partial_i \left[ \sqrt{-g} \, D \left( v^i - {\beta^i \over \alpha} \right) \right] = 0,\label{eq:ev_D} \\
\fl \partial_t \left( \sqrt{h} \, S_b \right) + \partial_i \left[ \sqrt{-g} \left( S_b \left( v^i - {\beta^i \over \alpha} \right) + P \, h^i{}_b
 - \frac{1}{W^2}\left(B^iB_b - \frac{1}{2} \, h^i{}_b  \, B^jB_j \right) 
   \right.\right.\nonumber\\
 \qquad\qquad \left.\left. - \frac{1}{2} \, B^j v_j \left(B^i v_b - \frac{1}{2} \, h^i{}_b  \, B^j v_j\right) \right) \right] \nonumber\\ 
\qquad\qquad\qquad  = \sqrt{-g} \, \left[ \, {^{3}{\Gamma}}_{ab}^i \left( \perp\! T \right)^a{}_i + {1\over\alpha} S_a \partial_b \beta^a - {1\over\alpha} \partial_b \alpha \, E \right], \\
\fl \partial_t \left( \sqrt{h} \, \tau \right) + \partial_i \left[ \sqrt{-g} \left( S^i - \frac{\beta^i}{\alpha} \, \tau - v^i D \right) \right] = \sqrt{-g} \, \left[ \left( \perp\! T \right)^{ab} \, K_{ab} - \frac{1}{\alpha} 
\, S^a \partial_a \alpha \right],\\
\fl \partial_t \left( \sqrt{h} \, B^b \right)  + \partial_i \left[ \sqrt{-g} \, \left( B^b \left( v^i - \frac{\beta^i}{\alpha} \right) 
  - B^i \left( v^b - \frac{\beta^b}{\alpha} \right) \right) \right] = 0 , 
\label{eq:ev_B} \\
\fl\frac{1}{\sqrt{h}} \, \partial_i  \left(\sqrt{h} \, B^i \right) = 0 
\label{eq:sol_const} ,
\end{eqnarray}
where the quantity $\bigl( \perp\! T \bigr)^i{}_b$ is the spatial projection 
of the stress tensor given in terms of the matter fields by 
\begin{eqnarray} 
\bigl( \perp\! T \bigr)^i{}_b 
  &=&   v^i S_b  
    + P \cdot h^i{}_b 
    - {1\over W^2} \, \Bigl[ B^i B_b - {1 \over 2} h^i{}_b \cdot B^2 \Bigr] 
                   \nonumber\\
   &&\qquad
    - \bigl( B^jv_j \bigr) \Bigl[ B^i v_b - {1\over2} h^i{}_b \cdot 
      \bigl( B^jv_j \bigr) \Bigr]  .  
\end{eqnarray} 
In the above, we work from a set of ``primitive" variables
${\bbw} = (\rho_0 , v^i , P , B^j)^{\rm T}$ consisting of the energy density,
$\rho_0$, the components of the coordinate velocity of the fluid,
$v^i$, the fluid pressure, $P$, and the magnetic field in the fluid frame,
$B^j$.
From these, we define a set of conservative variables,
${\bbu} = (D, S_b, \tau , B^j)^{\rm T}$ where the relativistic density $D$,
momentum $S_b$, and energy $E = \tau + D$ are given in terms of the primitive
variables by
\begin{eqnarray}
D &=& W \rho_0,\\
S_i &=& \left[h_e W^2 + B^2\right]v_i - \left(B^j v_j\right)B_i,\\
\tau &=& h_e W^2 + B^2 - P - \frac{1}{2}\left[\left(B^j v_j\right)^2
   + \frac{B^2}{W^2}\right] - W\rho_0,
\end{eqnarray}
and $B^j$ remains unchanged.  We have also defined $B^2 = B_i B^i$,
the fluid enthalpy $h_e = \rho_0 ( 1 + \epsilon ) + P$ with $\epsilon$ the
fluid's internal energy, and the Lorentz factor $W = (1-v_i v^i)^{-1/2}$.
Note that  spatial indices are lowered and raised by the 
3-metric $h_{ij}$ and its inverse.
Finally, to close the system, we assume a $\Gamma$-law equation of state
\begin{equation}
P = (\Gamma-1)\, \rho_0\epsilon,
\end{equation}
where $\Gamma$ is the usual adiabatic index.  Note that while Eqns.~\eref{eq:ev_D}--\eref{eq:ev_B} are indeed in balance law form, Eq.~\eref{eq:sol_const}
is not.  This last equation, of course, is the solenoidal constraint and
must be dealt with separately.   

One of the benefits of our chosen numerical scheme is that it does not 
require the full spectral decomposition of the system of evolution equations.  
However, we do find it useful to have some information about the 
possible speeds of some of the waves in our system.  This information 
comes by solving for the eigenvalues of the Jacobian matrix,
${\cal J\,}^k$, associated with the flux, ${\bbf\,}^k(\bbu)$, in the $k$
direction where    
\begin{equation} 
{\cal J}\,^k = \frac{\partial \bbf\,^k(\bbu)}{\partial \bbu} .  
\end{equation} 
On doing this for our system for the $k$ direction, say, one gets the general 
relativistic 
generalization~\cite{Anton:2005gi} of the seven wave speeds of  
flat space MHD~\cite{Anile,Komissarov:1999,DelZanna2002rv,Leismann:2005}.
These include the entropy wave and two Alfv{\'e}n waves,   
\begin{eqnarray}
{^{\rm ent}\lambda}^k & = & \alpha v^k - \beta^k \\
{^{\rm A}\lambda}^k_{\pm} & = & \alpha v^k - \beta^k - {B^k \over h_e W^2 + B^2} \left[ B^j v_j \pm \Bigl( h_e (B^j v_j)^2 + {B^2 \over W^2} \Bigr)^{1/2} \right] 
\end{eqnarray}
and the four (``fast" and ``slow") magnetosonic waves, 
${^{\rm f,s}\lambda}^k_{\pm}$, that are the zeros of the fourth 
order polynomial
\begin{eqnarray}
\fl 0 = h_e W^4 (1 - c_s^2) \bigl( \alpha v^k - \beta^k - \lambda^k \bigr)^4 \nonumber \\
\fl \qquad + \left[ \bigl( \beta^k + \lambda^k \bigr)^2 - \alpha^2 h^{kk} \right] \cdot 
\Bigl[ \bigl( \alpha v^k - \beta^k - \lambda^k \bigr)^2 \bigl( h_e W^2 c_s^2 + B^2 + W^2 ( B^j v_j )^2 \bigr) \Bigr. \nonumber \\ 
\Bigl. \qquad\qquad\qquad  
- \, c_s^2 \Bigl( W (B^j v_j) \bigl( \alpha v^k - \beta^k - \lambda^k \bigr) + \alpha {B^k \over W} \Bigr)^2 \Bigr] 
\label{eq:quartic}
\end{eqnarray}
where $c_s$ is the local sound speed and is given by 
\begin{equation}
h_e c_s^2 = \rho_0 \, {\partial P \over \partial \rho_0} + {P \over \rho_0} \, {\partial P \over \partial \epsilon} .  
\end{equation}
We solve \eref{eq:quartic} numerically using the DRTEQ4 routine from the
publicly available CERN Program Library.  The roots from DRTEQ4 are
then refined using a Newton-Raphson solver.
 
Finally, we note that the MHD equations are written in terms of both the 
conservative and primitive
variables.  The transformation from conservative variables to primitive 
variables is transcendental, requiring the solution of a single transcendental,
nonlinear equation, and is outlined in~\cite{Neilsen2005}.

%
%
\section{Numerical approach}
\label{sec:numerical}

This section briefly summarizes the numerical scheme we use to solve the
relativistic MHD equations. The fluid equations are solved with the
Convex Essentially Non-Oscillatory (CENO) scheme.  CENO is based on a 
finite difference discretization, which simplifies the coupling 
to the Einstein equations with AMR.  Hyperbolic divergence cleaning controls
growth of error in the solenoidal constraint, and gives some flexibility
in choosing other components of the numerical algorithm with AMR.

\subsection{CENO}
\label{subsec:methods}

The CENO scheme was developed by Liu and Osher~\cite{LiuOsher}
to efficiently solve equations in balance law form
\begin{equation}
\partial_t\bbu  + \partial_k\bbf\,^k(\bbu) = \bbs(\bbu),
\label{eq:balance2}
\end{equation}
where $\bbu$ is a state vector, $\bbf\,^k$ are flux functions, and $\bbs$ source terms.
We use the modification of CENO for relativistic fluids of Del Zanna and 
Bucciantini~\cite{DelZanna:2002qr}.
\Eref{eq:balance2} is solved using the method of lines.  The semi-discrete 
form in one dimension is
\begin{equation}
\frac{d \bbu_i}{dt} 
 = - \frac{\skew 8\hat\bbf_{i+1/2} - \skew 8\hat\bbf_{i-1/2}}{\triangle x}
   + \bbs(\bbu_i),
\label{eq:sd_eq}
\end{equation}
where $\skew 8\hat\bbf$ is a consistent numerical flux.   We use both
the Lax--Friedrichs flux and the HLL flux for the numerical flux.  The
Lax--Friedrichs flux is
\begin{equation}
{\bbf}\,^{\rm LF}_{i+1/2} 
    = \frac{1}{2}\left[ \bbf(\bbu^L_{i+1/2}) + \bbf(\bbu^R_{i+1/2})
          - (\bbu^R_{i+1/2} - \bbu^L_{i+1/2})\right],
\end{equation}
where $\bbu^L_{i+1/2}$ and $\bbu^R_{i+1/2}$ are the left and right
reconstructed states at $x_{i+1/2}$.  The HLL flux~\cite{Harten} is a 
central-upwind flux that 
uses the maximum characteristic velocities for both left- and
right-moving waves, $\lambda_\ell$ and $\lambda_r$, respectively,
\begin{equation}
{\bbf}\,^{\rm HLL} = \frac{\lambda^+_r \bbf(\bbu^\ell) 
         - \lambda^-_\ell\bbf(\bbu^r)
         + \lambda^+_r \lambda^-_\ell(\bbu^r - \bbu^\ell)}%
         {\lambda^+_r - \lambda^-_\ell},
\end{equation}
where
\begin{eqnarray}
\lambda^-_\ell &=& \min(0,\lambda_\ell)\\
\lambda^+_r &=& \max(0,\lambda_r).
\end{eqnarray}
For highly relativistic flows the
Lax--Friedrichs flux gives results very similar to the HLL flux; 
the maximum characteristic velocities approach the speed of light.

The point-valued fluxes ${\bbf}_{i+1/2}$ are then converted into
consistent numerical fluxes, $\skew 8\hat\bbf_{i+1/2}$.  For a second
order scheme, $\skew 8\hat\bbf_{i+1/2}={\bbf}_{i+1/2}$, while the correction
for the third-order scheme  is~\cite{ShuOsherI}
\begin{equation}
\skew 8\hat \bbf_{i+1/2}= \left( 1 - \frac{1}{24}{\cal D}^{(2)}\right)
         \bbf_{i+1/2},
\end{equation}
where ${\cal D}^{(2)}$ is a non-oscillatory second-order difference operator.
The operator used in this work is specified in~\cite{Neilsen2005}.

The accuracy of the overall numerical scheme is determined
by the reconstruction of the fluid states $\bbu^L_{i+1/2}$ and 
$\bbu^R_{i+1/2}$ from the solution known at grid points, i.e., the solution
at $x_{i-p}, \ldots, x_i, \ldots, x_{i+q}$ for integer $p$ and $q$.
Linear and quadratic reconstructions
discussed below lead to second and third order methods, respectively,
for smooth solutions.  As is commonly done in relativistic fluid dynamics,
we reconstruct the primitive variables rather than the conservative
variables.  This is because the conservative to primitive variable
transformation is transcendental, and computationally rather expensive.

The reconstruction is performed hierarchically, meaning that a reconstruction
of order $n$ is calculated from the reconstruction of order $n-1$ using a
general algorithm.  This allows one to construct schemes of formally very 
high order.  Thus, we first obtain a linear reconstruction, $L_i$, which 
is then used to create the second order reconstruction.  $L_i$ is defined on 
the domain $[x_{i-1/2},x_{i+1/2}]$ as
\begin{equation}
L_i(x) = u_i + u'_i(x-x_i),
\end{equation}
where $u'_i$ is the limited slope
\begin{equation}
u'_i = \mbox{minmod}(D_- u_i, D_+ u_i).
\end{equation}
Here we have defined one-sided and centered difference operators as
\begin{equation}
D_{\pm} u_i = \pm \frac{u_{i\pm 1} - u_i}{\triangle x},\qquad
D_{0} u_i   = \frac{u_{i+1} - u_{i-1}}{2\triangle x},
\end{equation}
and the minmod limiter is
\begin{equation}
\mbox{minmod}(a_1, a_2, \cdots) = 
\left\{
\begin{array}{ll}
\min\{a_k\} & \hbox{if all $a_k > 0$,}\\
\max\{a_k\} & \hbox{if all $a_k < 0$,}\\
0 & \hbox{otherwise.}
\end{array}\right.
\label{eq:minmod}
\end{equation}
The first-order reconstruction, $L_i(x)$, is thus equivalent to the 
linear TVD reconstruction.

A second order reconstruction proceeds by constructing three candidate
quadratic functions, $Q^k_i(x)$, which are then compared to $L_i(x)$.
When the solution is smooth, one of the quadratic functions is chosen
for the reconstruction.  Near discontinuities, however, the linear
reconstruction is retained, thus giving solutions similar to TVD schemes
for discontinuous solutions.
The three candidate quadratic functions are 
\begin{equation}
Q^k_{i}(x) = u_{i+k} + D_0 u_{i+k} (x - x_{i+k})
            + \frac{1}{2}D_+D_- u_{i+k} (x - x_{i+k})^2 ,
\end{equation}
with $k=-1, 0, 1$.  Weighted differences with respect to $L_i(x)$
are then calculated
\begin{equation}
d^k(x) = \alpha^k \left( Q^k_i(x) - L_i(x)\right).
\label{eq:eno_differences}
\end{equation}
The weights $\alpha^k$ are chosen to bias the reconstruction towards the 
centered polynomial:
$\alpha^0 = 0.7$, and $\alpha^{-1} = \alpha^1 = 1$. 
When the differences $d^k$ all have the same sign, we choose the $Q^k_i(x)$
for which $d^k$ has the smallest magnitude.
When the $d^k(x)$ have differing signs, we revert to the first order 
reconstruction.

Finally, the semi-discrete equations are integrated with the optimal 
third-order Runge--Kutta that preserves the TVD condition~\cite{ShuOsherI}
\begin{eqnarray}
\bbu^{(1)} &=& \bbu^{n} + \triangle t L(\bbu^n),\nonumber\\
\bbu^{(2)} &=& \frac{3}{4}\bbu^{n} + \frac{1}{4}\bbu^{(1)} 
           + \frac{1}{4}\triangle t L(\bbu^{(1)}),\\
\bbu^{n+1} &=& \frac{1}{3}\bbu^{n} + \frac{2}{3}\bbu^{(2)} 
           + \frac{2}{3}\triangle t L(\bbu^{(2)}).\nonumber
\end{eqnarray}

\subsection{Hyperbolic Divergence Cleaning}
\label{subsec:hdc}

The time evolution of the magnetic field is governed by \eref{eq:ev_B}
above. However, $\bf B$ is also subject to the solenoidal constraint
$\nabla \cdot {\bf B}=0$. 
The continuum evolution equations preserve this constraint, although
it may be violated in numerical evolutions. These
violations can lead to unphysical
numerical solutions~\cite{BrackbillBarnes,Brackbill}.  Some differencing
schemes for the Maxwell equations and MHD are designed such that
a particular discretization of the solenoidal constraint 
is satisfied to machine precision.
These constrained transport methods, naturally, do not give solutions
that exactly satisfy the continuum constraint, and the magnitude
of the constraint error can be estimated by using an independent
discrete divergence operator.  Constrained transport methods for
classical MHD have been used with 
AMR~\cite{Balsara2001b,TothRoe,LiLi,Balsara2004}.

Divergence cleaning methods are an alternative approach to constrained
transport, and allow some flexibility in designing the numerical
algorithm.  Elliptic divergence cleaning methods require the
solution of a Poisson equation (either explicitly, or implicitly in
Fourier space), and some common implementations have been reviewed for 
classical MHD by T\'oth~\cite{Toth} and Balsara and Kim~\cite{BalsaraKim}.  
T\'oth reports favorably on divergence cleaning, while Balsara and Kim
argue that constrained transport performs better for a wider class of
problems.  Hyperbolic divergence cleaning
is quite efficient, easy to implement, and usually gives
good results~\cite{Dedner2002}.  
A new field $\psi$ is added to the
equations and coupled to the evolution equations for $\bf B$.  The field $\psi$
acts as a generalized Lagrange multiplier, similar to the $\lambda$-system
used in solving the Einstein equations~\cite{Brodbeck1999}.
Having some freedom in choosing the equation for $\psi$,
we choose
\begin{eqnarray}
\label{eq:newB}
\partial_t B^b + \partial_i\left( B^b v^i - B^iv^b\right) 
           + h^{bj}\partial_j \psi &=& 0,\\
\qquad\qquad \frac{1}{c_h^2}\partial_t \psi + \frac{1}{c_p^2}\psi 
+ {\bf \nabla} \cdot {\bf B} &=& 0.
\label{eq:psi}
\end{eqnarray}
It can be shown that $\psi$ satisfies the telegraph equation, whose
solutions are damped, traveling waves.  The parameters $c_h$ and $c_p$ 
control the speed and damping of the constraint propagation, respectively.  
We use $c_h = 1$ and $c_p \in \left[1,12\right]$ in the tests examined here.
Using larger values of $c_h$ requires smaller Courant factors and did 
not change results significantly in~\cite{Neilsen2005}.  In contrast, the 
optimal damping factor, $c_p$, is related to the size of the initial shock 
discontinuity.  Generally, the larger the shock, the larger the
value necessary for $c_p$.  Finally, work is underway to develop constraint 
preserving boundary conditions consistent with hyperbolic divergence 
cleaning for the MHD equations~\cite{Palenzuela}.

\subsection{Adaptive Mesh Refinement}
\label{subsec:amr_methods}

AMR provides the ability to add grid refinement where and when needed. This
need is determined by some refinement criterion. At any given level of
resolution, points which meet this criterion are flagged and a new, finer
level is created which includes all such flagged points. Similarly, when
no points are flagged, the level is removed. For the tests presented here,
the maximum number of levels of refinement was limited to two. In other words,
our runs have grids with three different resolutions.

The fine and coarse grids communicate in AMR through prolongation
and restriction.  Fine grids are created by interpolating the
solution from a parent grid (prolongation), and the fine grid
solution is communicated to coarser grids through restriction.
In prolongation we interpolate the conservative variables onto
finer grids using third-order WENO 
interpolation~\cite{SebastianShu,Neilsen2005}.  This
interpolation scheme is designed to work well with discontinuous
functions by adjusting the interpolation stencil to the local smoothness
of the function. This avoids oscillations near discontinuities, which often
cause primitive variable solvers for relativistic fluids to fail.  For
restriction on vertex centered grids, 
the fine grid values are copied directly to the coarse grid (direct injection).
If a point on a vertex centered coarse grid is also found on two or more
finer grids, the restriction operation averages the values on the
finer grids for the solution at the coarser grid point.

The distributed AMR infrastructure that we employ is {\sc had}.  
{\sc had} is a F77 based Berger-Oliger~\cite{Berger} type
AMR package presented in~\cite{Liebling} using the message passing
interface (MPI) for distributed parallelism.  {\sc had} has a modular design, 
allowing one to solve different sets of equations with 
the same computational infrastructure.
Unlike many other publicly available AMR toolkits,
including~\cite{SAMRAI,Chombo,Paramesh,AMROC,Boxlib}, {\sc had} is natively
vertex centered.
This can be advantageous in numerical relativity, as
many difference schemes for the Einstein equations are vertex centered.
Support for cell centered grids in {\sc had} is also available.
{\sc had} supports subcycling of grids in time for full
space-time AMR and can in principle accommodate arbitrary
orders of accuracy in both space and time.  An example has been
shown using third order accurate AMR
simulations~\cite{Lehner2006}.  The {\sc had} clustering algorithm is
Berger-Rigoutsos~\cite{Rigoutsos}; the load balancing algorithm is
the least loaded scheme~\cite{Rendleman}.  

Considerable flexibility is provided in developing a refinement 
criterion for the {\sc had} infrastructure, and many were explored in 
conjunction with the numerical tests presented here, such as refining on
gradients in the density, pressure or the magnetic field.
All criteria resulted in similar adaptive mesh hierarchies for the 
relativistic rotor and spherical blast wave tests;
consequently, those tests center the refinement on the evolving shock front.
A shadow hierarchy~\cite{Pretorius} has recently been added
to {\sc had} for truncation error estimation and was used as the 
refinement criterion in the spherical
accretion tests of a fluid falling onto a Schwarzschild black hole.

%
%
\section{Numerical Results}
\label{sec:nr}

In this section we examine three relativistic MHD test problems 
and one accretion test problem
using AMR. The MHD test problems are selected
because their solutions have very sharp features that require high resolutions.
The accretion test problem demonstrates the AMR capabilities with a curved space background
and an excision region.

The first problem is a one-dimensional blast wave problem introduced by 
Balsara~\cite{Balsara2001}.  We compare the unigrid shocktube results with 
the exact solution, and then compare unigrid results with the 
resolution-equivalent AMR results.  
The second and third problems are three-dimensional extensions of standard
two-dimensional tests:  a spherical blast wave and a spherical relativistic 
rotor~\cite{DelZanna2002rv,Shibata:2005gp,Neilsen2005}.
We present unigrid and AMR results for these test problems, and discuss the
effects of hyperbolic divergence cleaning.
The last problem is the accretion of a fluid onto a Schwarzschild black hole.
We numerically recover the steady state solution using AMR with a shadow 
hierarchy as the refinement criterion.
Finally, we present scaling results for parallel unigrid and AMR runs.
In the tests below we use the discrete $L_2$ norm 
\begin{eqnarray}
|| \, u \, ||_2 = 
     \left[ {1\over N-1} \, \sum_i^N \left(u_i\right)^2 \right]^{1/2}
\end{eqnarray}
where $u$ is a discrete function defined at $N$ locations, $u_i$.

\subsection{Riemann problem test}
\label{subsec:rpt}

Balsara introduced several test Riemann problems for 
relativistic MHD~\cite{Balsara2001},  and we choose here his third blast wave
problem for its very narrow features as a test of our AMR.  
The initial parameters for this Riemann problem are given 
in \tref{table:blastwave}.
The exact solution for this problem is given by Giacomazzo and 
Rezzolla~\cite{Giacomazzo2005jy}, and is plotted in the figures below for 
comparison.

\begin{table}
\begin{center}
{\bf Balsara blast wave parameters}
\begin{tabular}{c|c|c|c|c|c|c|c|c}
\hline
      &  $\rho_0$ & P    & $v^x$ & $v^y$ & $v^z$ & $B^x$ & $B^y$ & $B^z$ \\
  \hline
Left  &   1.0   & 1000 &  0.0  & 0.0   & 0.0   & 10.0  & 7.0   & 7.0  \\
Right &   1.0   & 0.1  &  0.0  & 0.0   & 0.0   & 10.0  & 0.7   & 0.7  
\end{tabular}
\caption{The initial parameters for the Balsara blast wave.  
The discontinuity is initially placed at $x=0$ and the system is evolved 
along the $x$ axis.  The adiabatic index is $\Gamma = 5/3$.  
The exact solution to this problem is given in \cite{Giacomazzo2005jy}.
The Courant factor used in the numerical solution is $0.2$.}
\label{table:blastwave}
\end{center}
\end{table}

\begin{figure}
\begin{tabular}{cc}
\epsfig{file=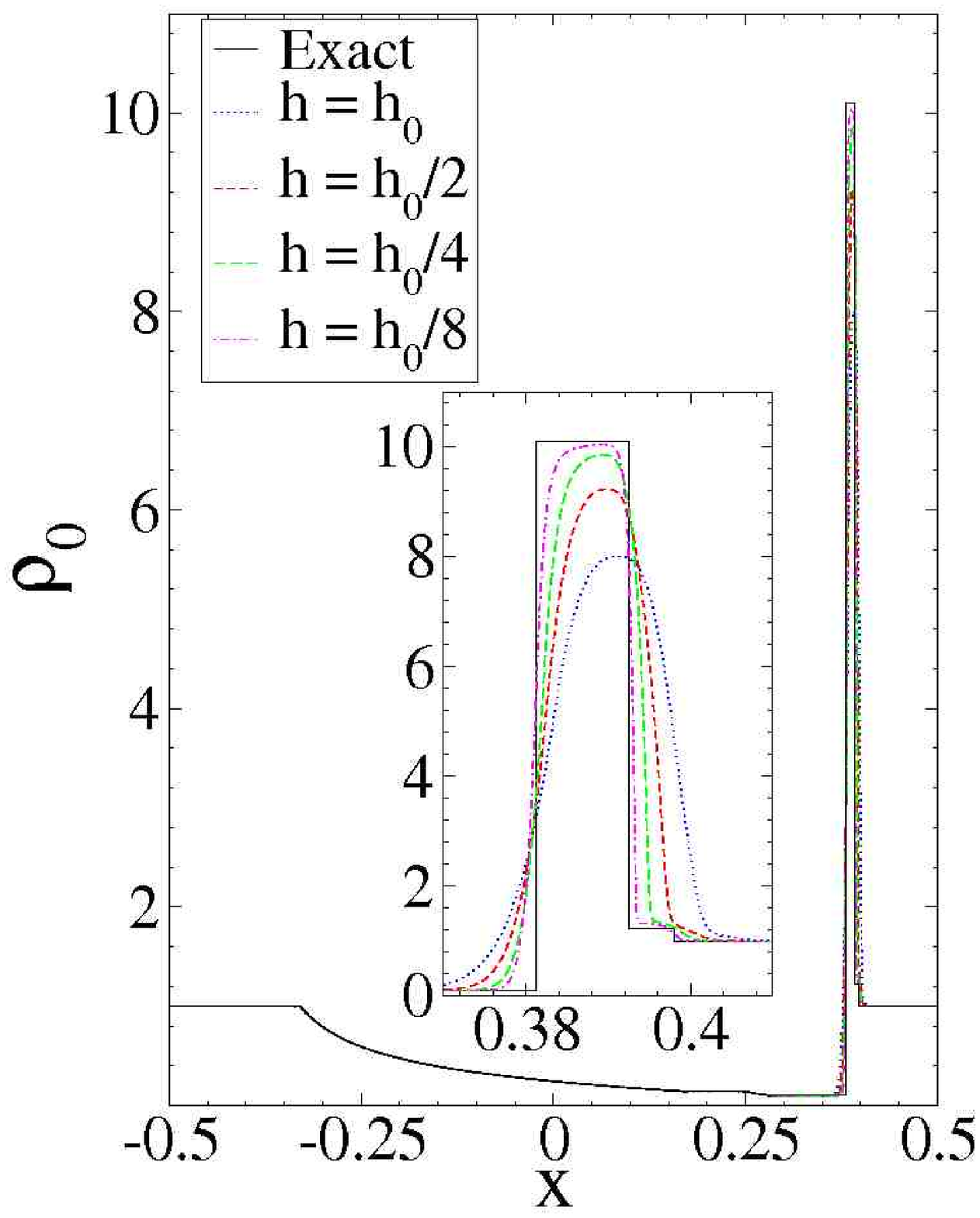,height=8.5cm} & \epsfig{file=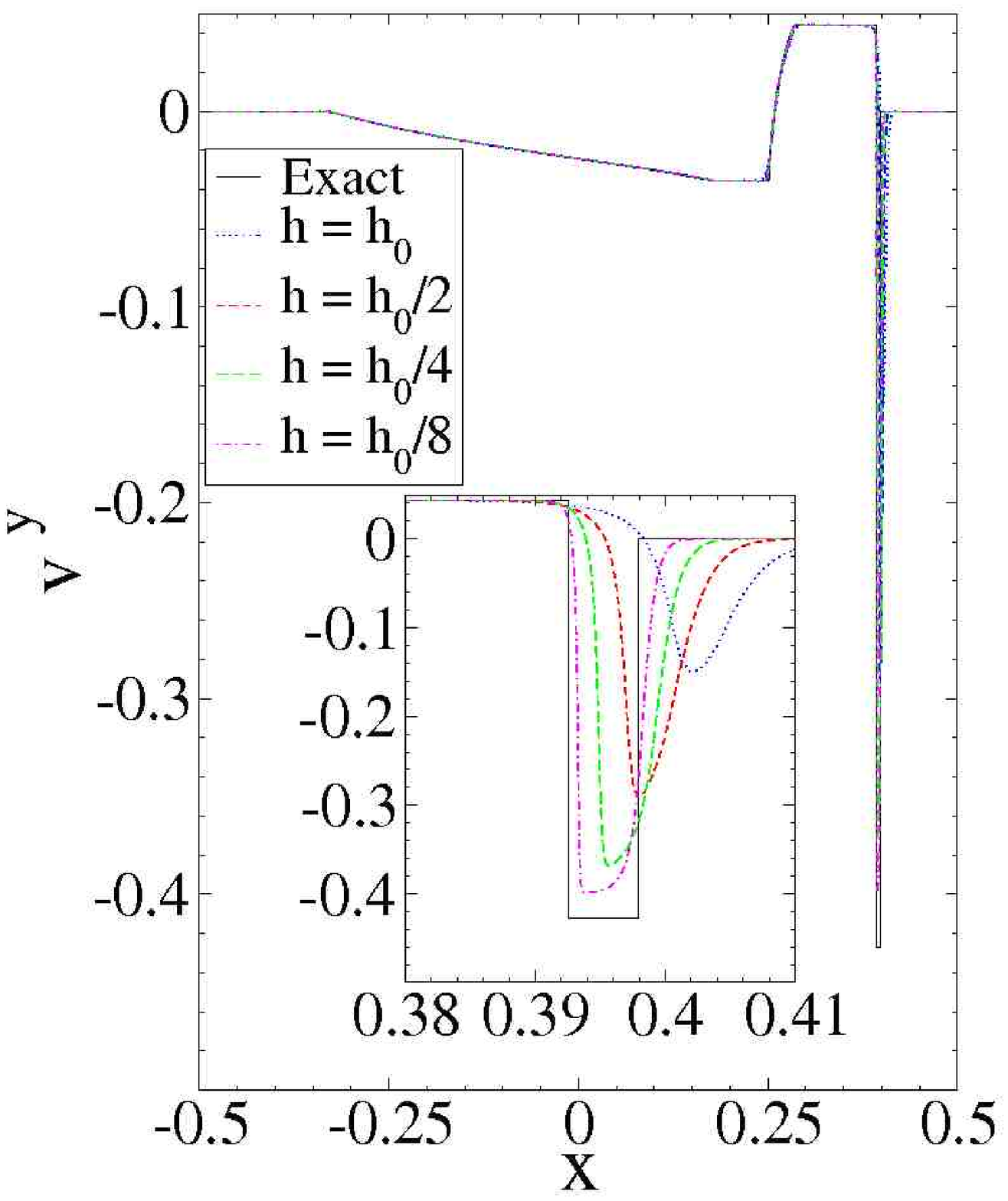,height=8.5cm} \\
{\bf (a)} & {\bf (b)} \\
\epsfig{file=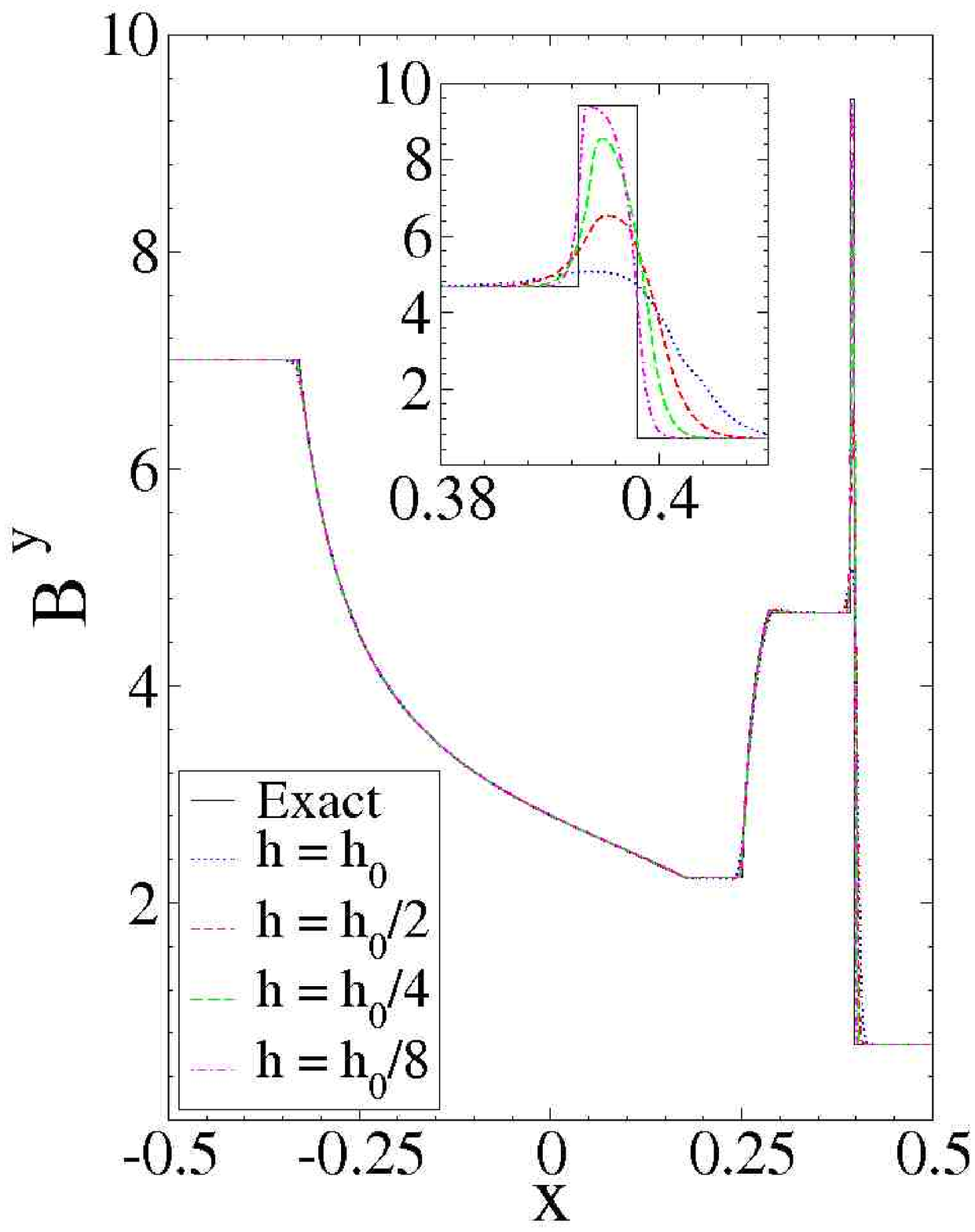,height=8.5cm}  & \epsfig{file=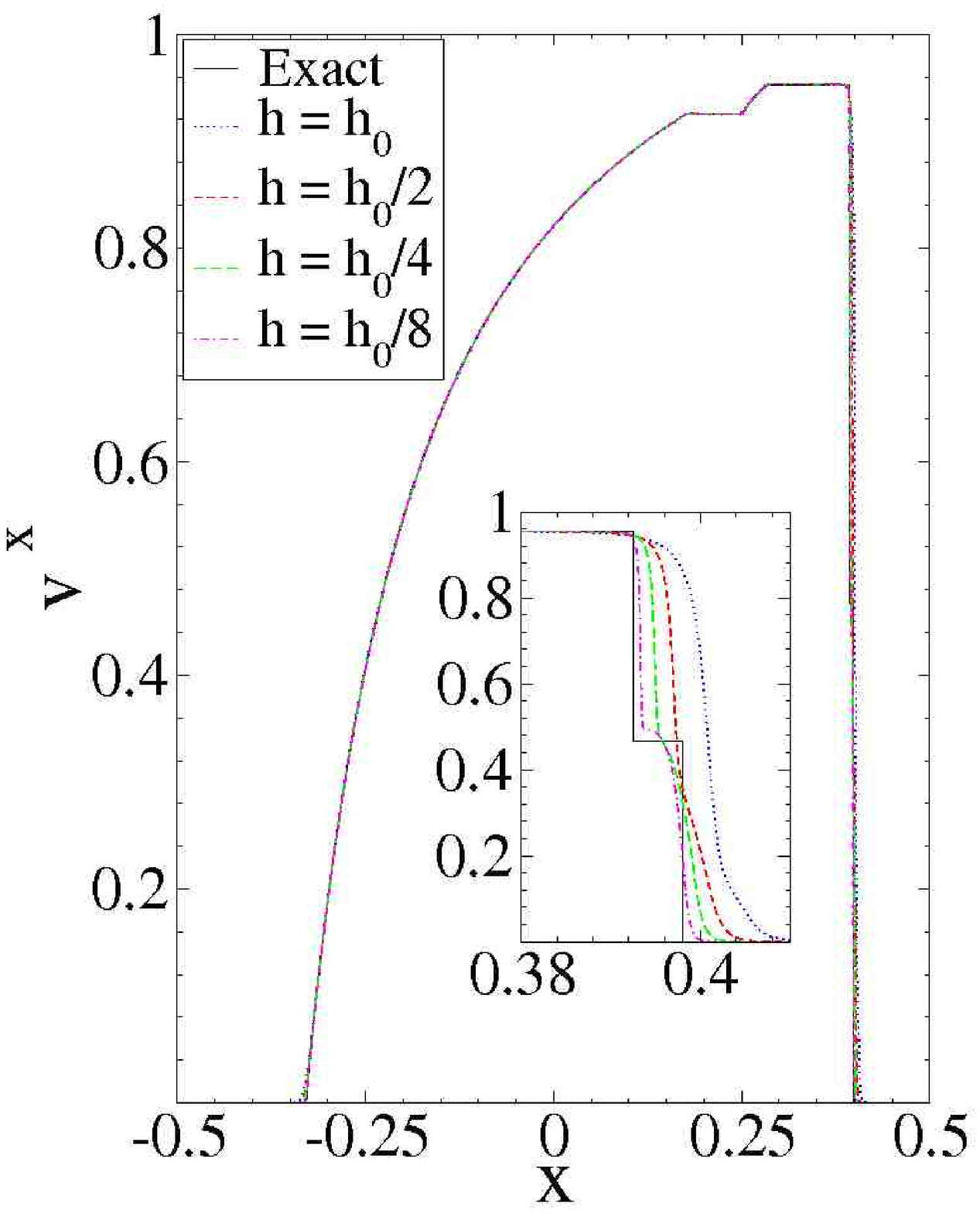,height=8.5cm} \\
{\bf (c)} & {\bf (d)} \\
\end{tabular}
\caption{Unigrid simulation results for the relativistic MHD Balsara blast 
wave test at time 0.4 showing $\rho_0$, $v^x$, $v^y$, and $B^y$.  
The $z$ components of $\bf B$ and $\bf v$ are identical to their 
respective $y$ components.  The simulations were performed
along the $x$ axis using four resolutions on a 
domain of $\left[-0.5,0.5\right]$. 
The base resolution was $h_0 = 6.25 \times 10^{-4}$.
The exact solution to this problem is found in~\cite{Giacomazzo2005jy}.  
This problem is an excellent candidate for AMR because of the high 
resolutions required to adequately resolve the different waves.}
\label{fig:unishock}
\end{figure}

The blast wave problem is implemented using the three-dimensional {\sc had} 
infrastructure for AMR and simulated along a coordinate axis.  
\Fref{fig:unishock} shows the blast wave with a strong initial pressure 
difference centered at $x=0$ evolved along the $x$-axis
at several unigrid resolutions.  The plots show $\rho_0$, $v^x$, $v^y$, 
and $B^y$ at time $t=0.4$.
The unigrid simulations were performed 
in the $x$ direction on a domain
of $\left[-0.5,0.5\right]$ with a Courant factor of $0.2$.

\begin{figure}
\begin{tabular}{cc}
\epsfig{file=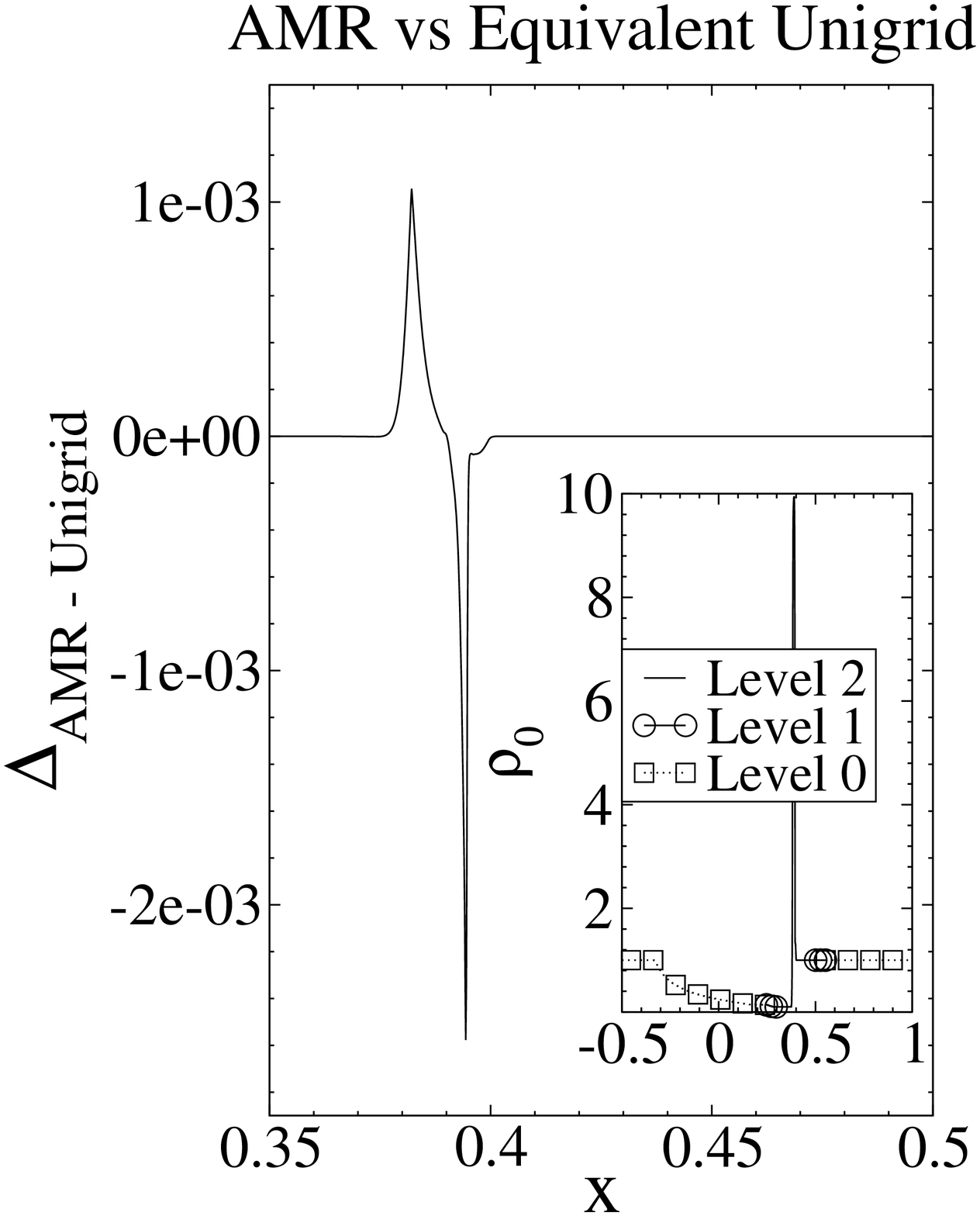,height=8.5cm} & \epsfig{file=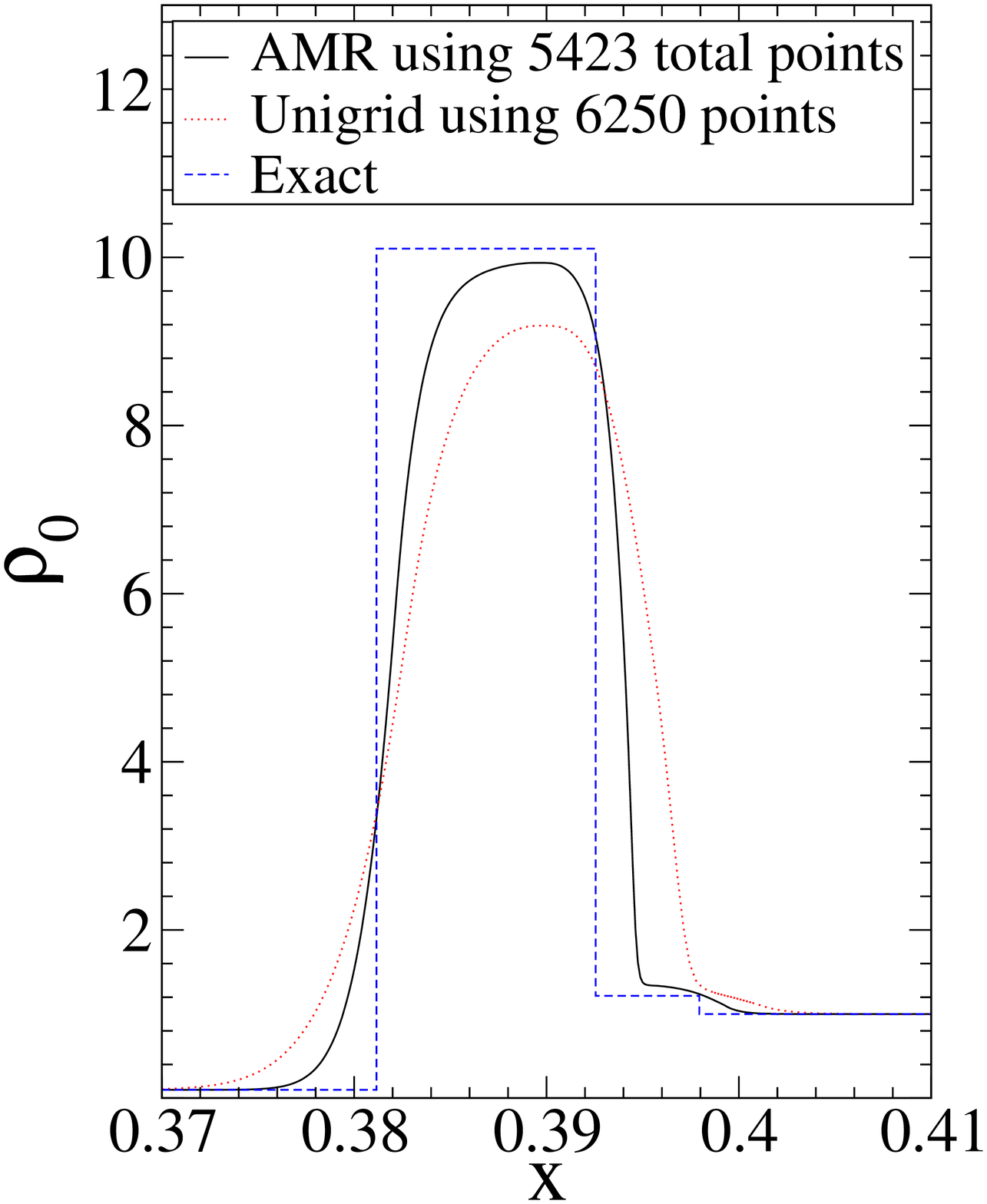,height=8.5cm} \\
{\bf (a)} & {\bf (b)} 
\end{tabular}
\caption{Adaptive mesh refinement results for the Balsara blast wave.
\Fref{fig:amrlayout}(a) shows the difference in $\rho_0$ between the finest mesh
in a two-level AMR simulation and the equivalent unigrid
simulation.  The finest mesh of the AMR simulation
has the same resolution as the unigrid case, $h = 1.25 \times 10^{-4}$.
Simulations were performed on a domain of $\left[-0.8,1.2\right]$.
The unigrid required 16000 points while the
AMR required 5423 points.  For reference, $\rho_0$ and the
layout of the AMR mesh hierarchy at $t=0.4$ are shown in the window inset.
\Fref{fig:amrlayout}(b) compares two solutions of roughly equal 
computational cost: one unigrid using 6250 points with resolution
$h = 3.2 \times 10^{-4}$ and the other AMR
using 5423 points with maximum resolution $h = 1.25 \times 10^{-4}$.  
The AMR simulation produces significantly more accurate
results for $\rho_0$ than the unigrid solution at the same computational cost.}
\label{fig:amrlayout}
\end{figure}

A series of two-level AMR simulations of the blast wave test were
conducted with refinement centered on the shock propagation.
\Fref{fig:amrlayout} compares a unigrid simulation with a two-level
AMR simulation where the resolution of the finest mesh in the AMR
hierarchy is the same as that in the unigrid simulation.  The AMR
capability to reproduce the resolution-equivalent unigrid result
depends on how well the refinement region tracks the shock throughout
the time history of the evolution.  In our tests we observe the AMR
blast wave simulations reproducing the resolution-equivalent unigrid
result to within $~0.1\%$.

\subsection{Spherical blast wave}
\label{subsec:sbw}

The spherical blast wave consists of a uniform fluid background with a small
spherical region where the pressure is $10^6$ times larger than the background.
The background pressure is $10^{-2}$, and $P=10^4$ inside a central sphere
of radius $0.08$.
This is the three-dimensional extension of the cylindrical blast wave
studied in~\cite{DelZanna2002rv,Shibata:2005gp,Neilsen2005}.
The parameters for the spherical blast wave are given in
\tref{table:sphblastwave}.

\begin{figure}
\begin{tabular}{cccc}
&With Cleaning && Difference \\
  \hline
$P$ & \includegraphics*[height=6.0cm]{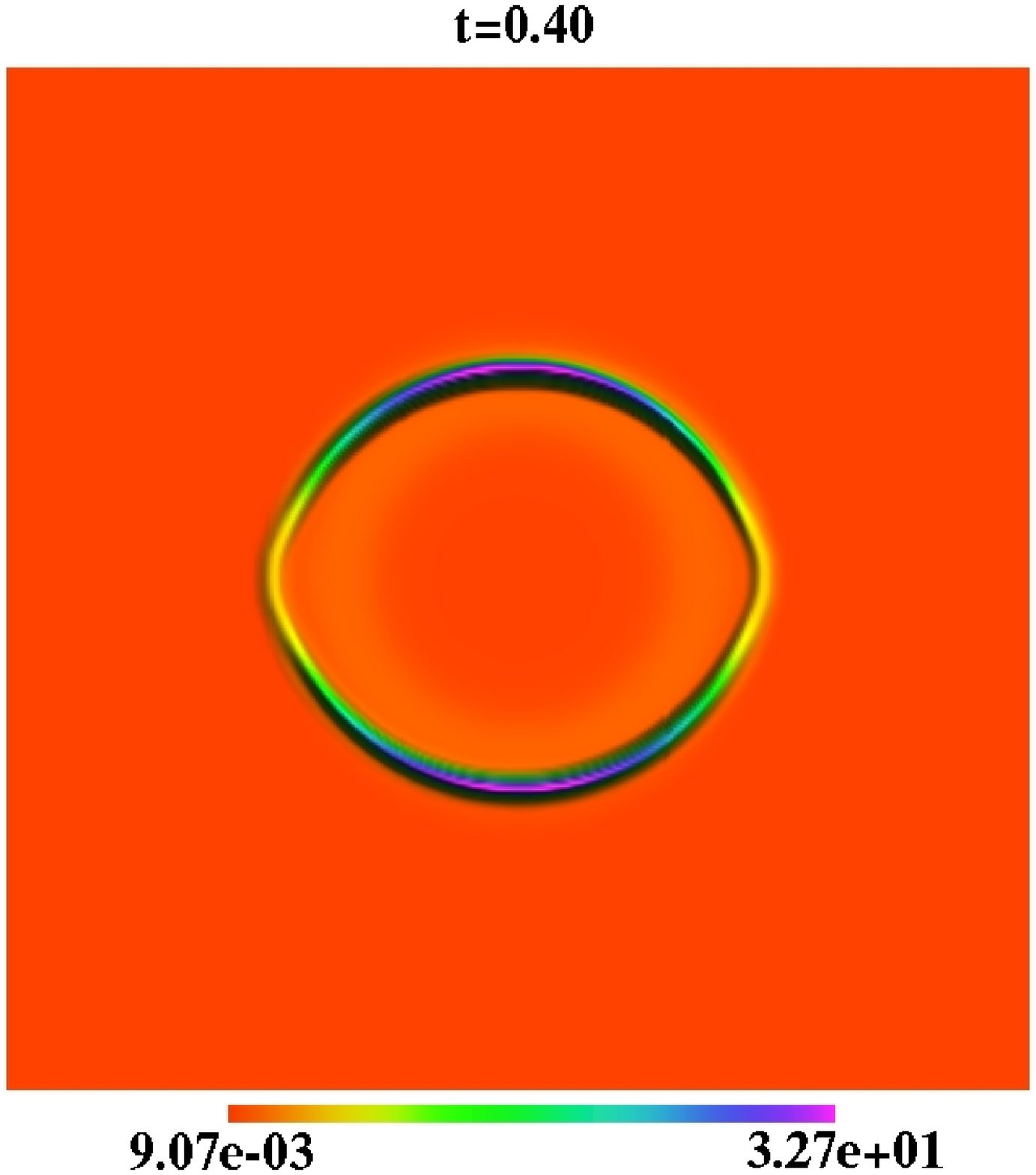}&
$\Delta P$& \includegraphics*[height=6.0cm]{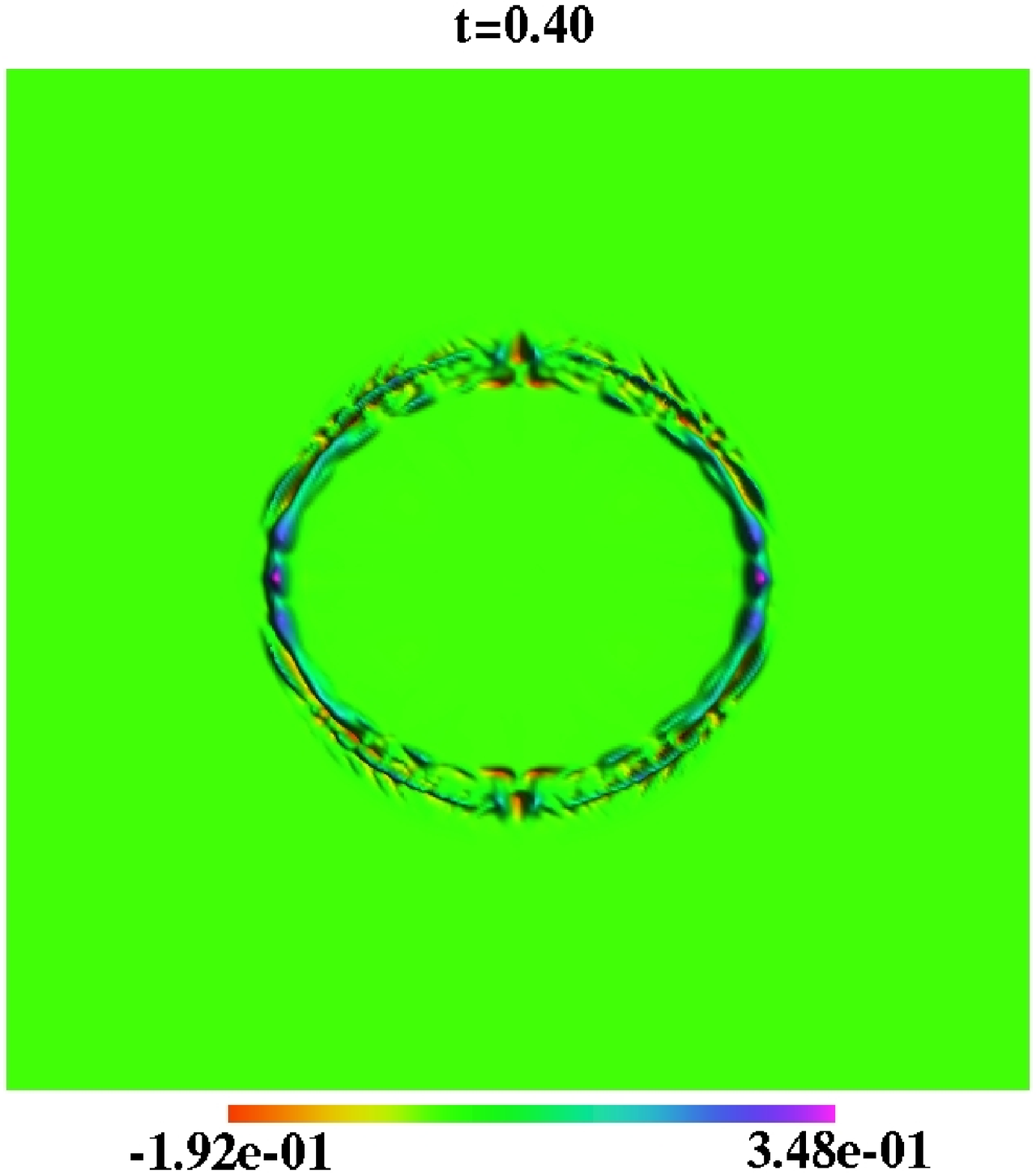}
\end{tabular}
\caption{Unigrid simulation results for the spherical blast wave at $t=0.4$,
showing a slice along the $z=0$ plane.  The $x$ axis is the horizontal 
direction.  The pressure found using hyperbolic divergence cleaning 
is shown on the left.
The difference between the pressure found with and without hyperbolic
divergence cleaning is shown on the right.  This 
gives an estimation of the relative errors that
arise in free evolutions.
The simulations were performed using a resolution of $h=0.006410$ on a domain 
of $\{x,y,z\}\in\left[ -1,1\right]$.}
\label{fig:2dsphshock}
\end{figure}

\begin{figure}
\begin{tabular}{cc}
\epsfig{file=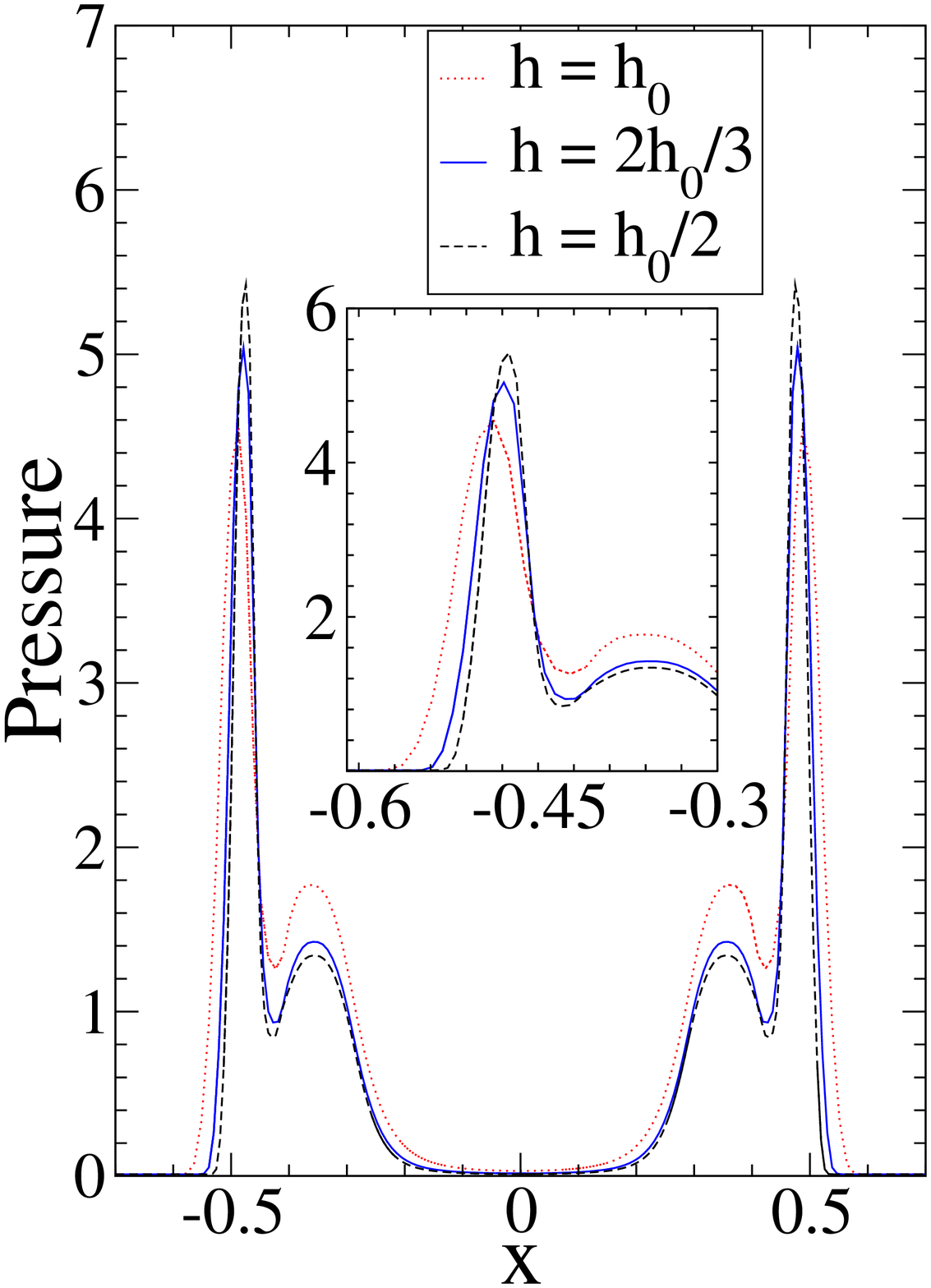,height=8.0cm} & \epsfig{file=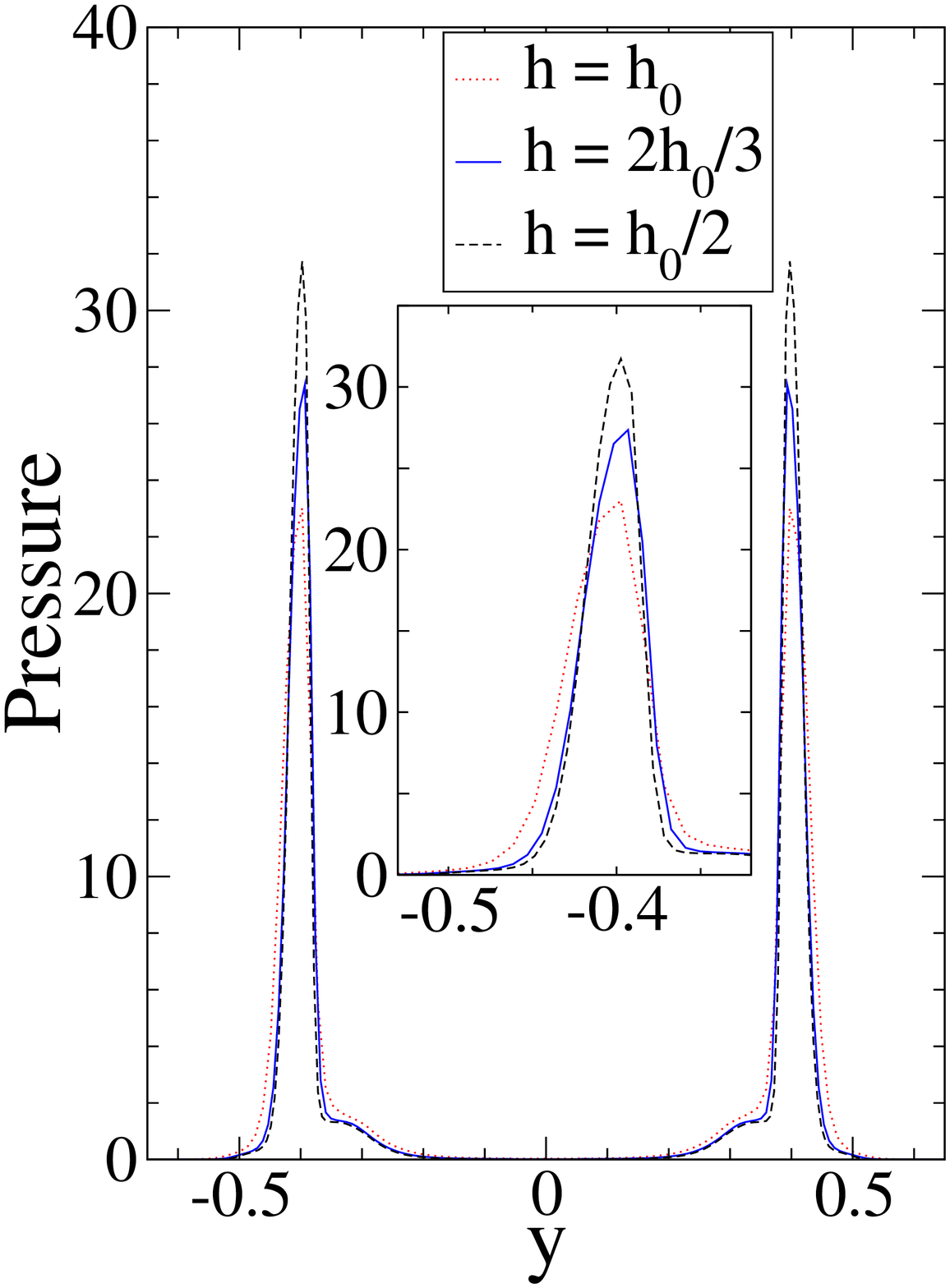,height=8.0cm} \\
{\bf (a)} & {\bf (b)}
\end{tabular}
\caption{The pressure for the spherical blast wave at $t=0.4$ shown at three 
resolutions along the $x$-axis (left frame) and the $y$-axis (right frame).
The window insets provide a closer view of the shock front amplitudes at the
different resolutions.  The base resolution is $h_0 = 0.0128$.}
\label{fig:convSphShock}
\end{figure}

\begin{figure}
\begin{center}
\epsfig{file=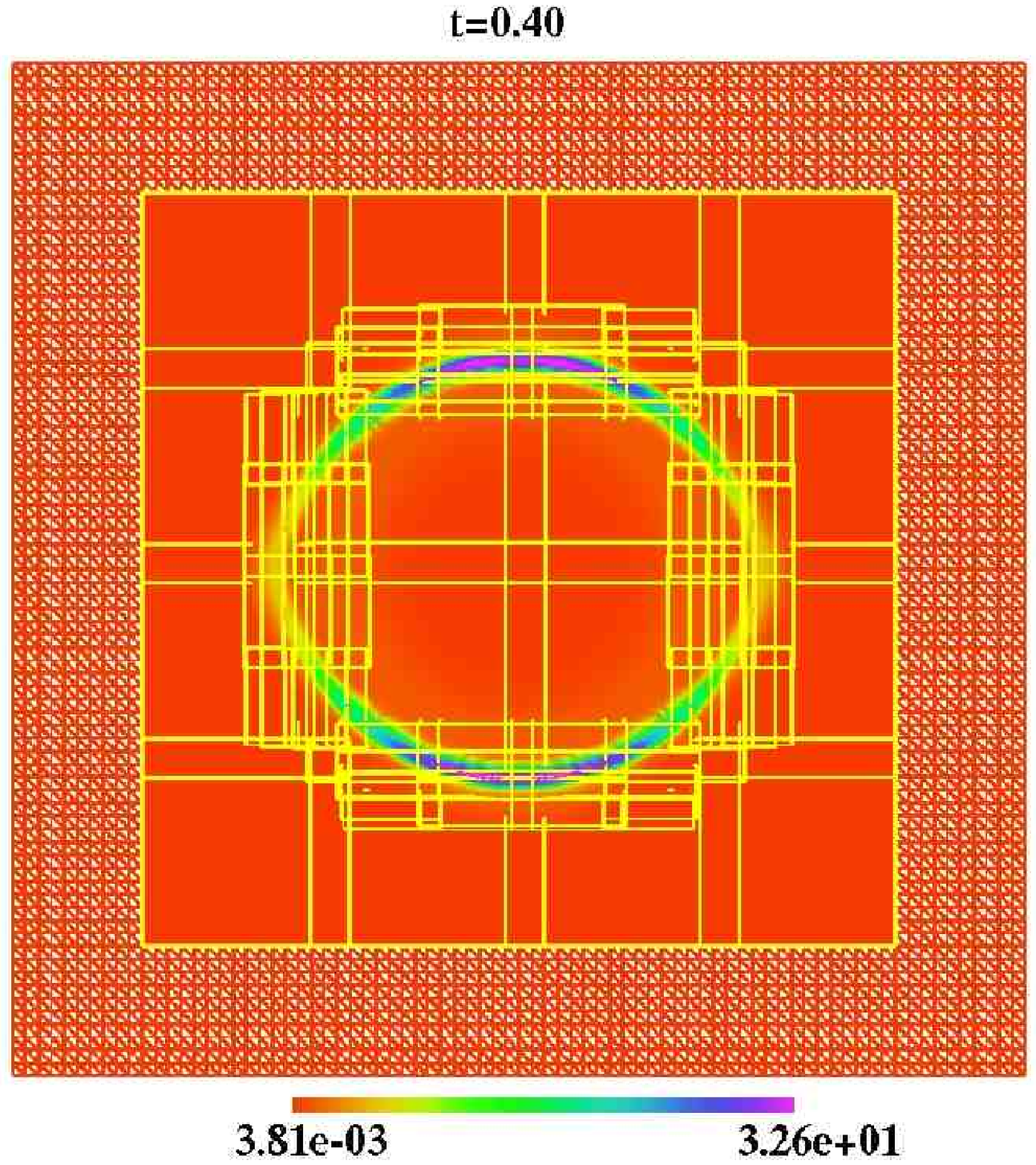,height=9.0cm} \\  
{\bf (a)} \\
\vspace{10pt}
\epsfig{file=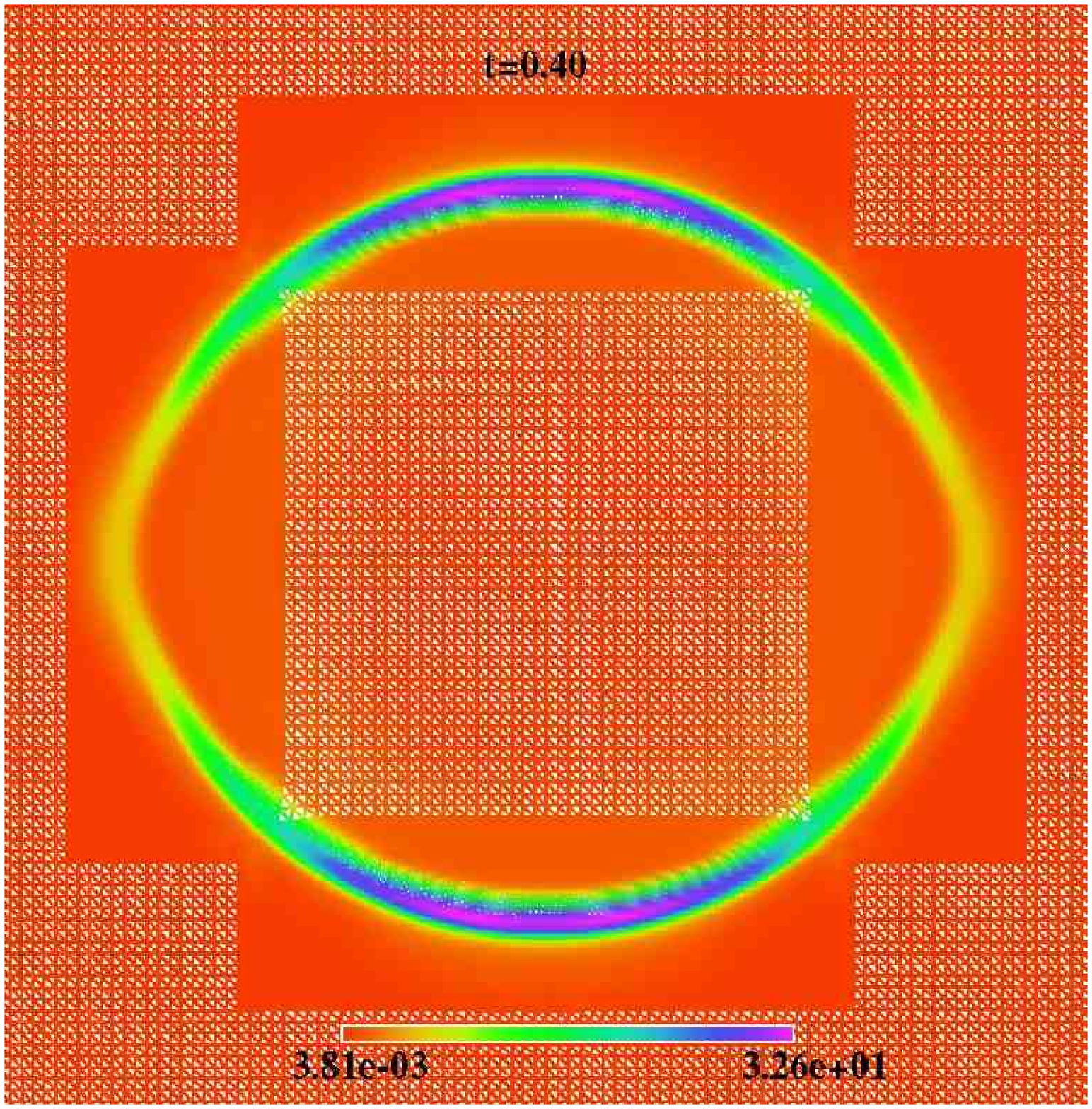,height=8.0cm} \\
{\bf (b)} \\
\end{center}
\caption{A wireframe slice along the $z=0$ plane of $P$
for the spherical blast wave using hyperbolic divergence
  cleaning and two levels of adaptive mesh refinement.  In each plot, 
the $x$ axis is the horizontal direction.
The top frame illustrates the domain decomposition of the system across
32 processors.  Only two separate resolutions are distinguishable,
but a third resolution becomes apparent in close-up, shown in the
the bottom frame.
The simulation reproduces the peak unigrid amplitude to within 0.4\%.
This AMR simulation is eight times faster than the equivalent unigrid 
simulation (shown in \fref{fig:2dsphshock}) and
was performed with a maximum resolution of 
$h=0.006410$ on a domain of $\{x,y,z\}\in\left[ -1,1\right]$ 
using 32 processors.}
\label{fig:amr-sphshock}
\end{figure}

We first calculate the solution on a single uniform grid, and then draw
comparisons to the AMR results.
\Fref{fig:2dsphshock} shows $z=0$ cuts of the uniform grid solution at
$t=0.4$, and \fref{fig:convSphShock} shows line plots of the pressure, $P$, 
along the $x$- and $y$-axes at three different resolutions.
Adaptive mesh refinement substantially improves performance for the
spherical blast wave while returning results nearly identical to
the unigrid result.  \Fref{fig:amr-sphshock} shows the resulting
mesh hierarchy and pressure at time $t=0.4$ for the spherical blast
wave in an AMR simulation with two levels of refinement.  This AMR
simulation uses hyperbolic divergence cleaning and is the AMR
equivalent of the divergence cleaning unigrid simulation in
\fref{fig:2dsphshock}.  The refinement criteria are set to center refinement
on the shock.  The relative simplicity of this solution---the outgoing shock 
is the dominant feature of the solution--allows
many refinement criteria to produce similar mesh hierarchies.
The AMR simulation in
\fref{fig:amr-sphshock} was performed on 32 processors and was a factor
of eight times faster than the equivalent unigrid simulation.
Like the Balsara blast wave case examined in \ref{subsec:rpt},
adaptivity significantly reduces the computational overhead required
to adequately resolve the multiscale features that appear in the
simulation.

\begin{figure}
\begin{center}
\epsfig{file=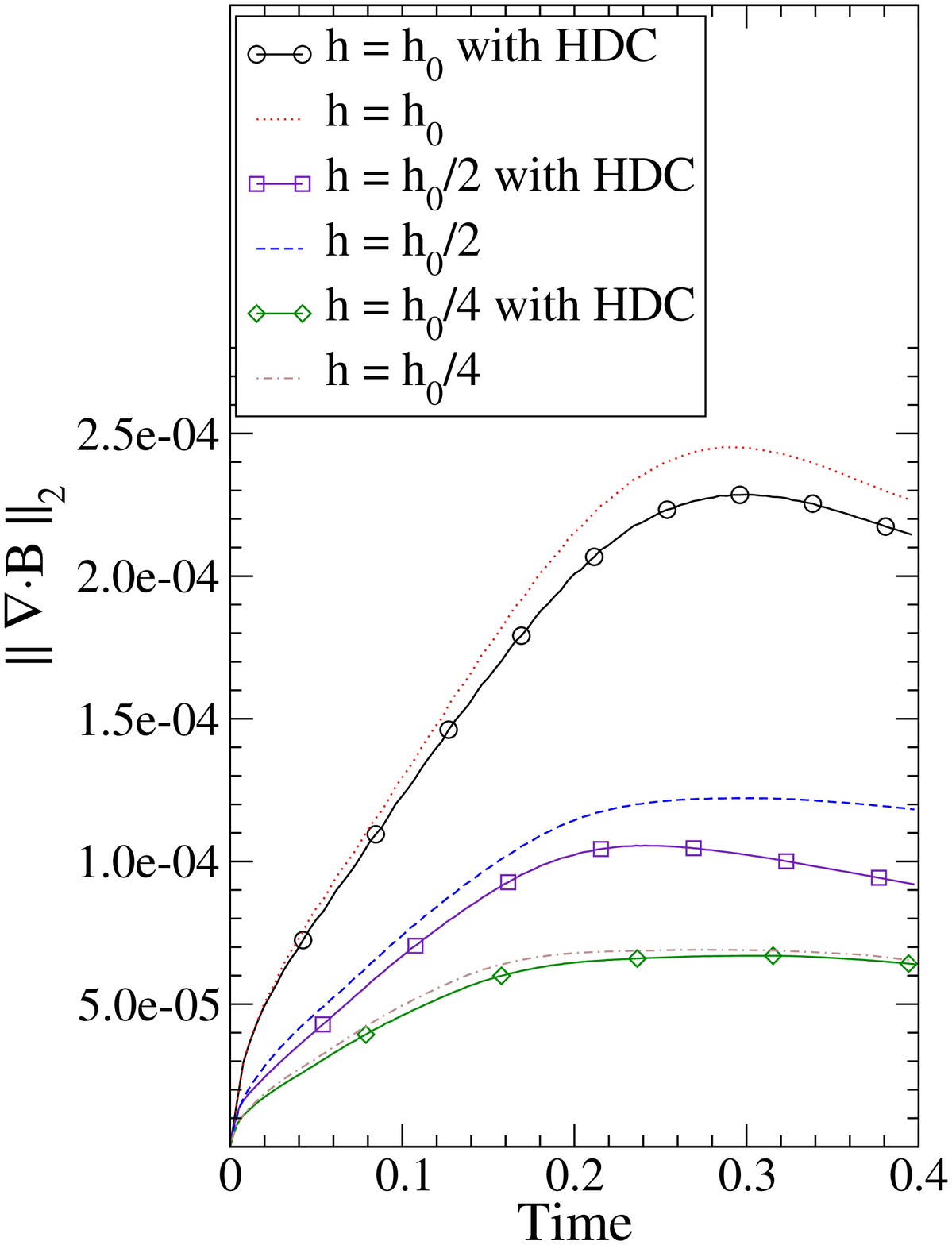,height=12.8cm}
\end{center}
\caption{The $L_2$ norm of ${\bf \nabla} \cdot {\bf B}$ as a function of 
time for the spherical blast wave at three resolutions for both free
evolutions and evolutions with hyperbolic divergence cleaning.
The base resolution is $h_0 = 0.025641$.
The divergence cleaning parameters are $c_h = 1$ and 
$c_p = 12$.}
\label{fig:sphshock-constraint}
\end{figure}

Finally, we monitor the violations of the solenoidal constraint during
both free evolutions and evolutions with hyperbolic divergence cleaning.
\Fref{fig:sphshock-constraint} shows the magnitude of the $L_2$ norm of $\nabla\cdot {\bf B}$
at three different resolutions with and without divergence cleaning.
There are some subtleties in interpreting $L_2$ norms of the 
constraint violation, which arise primarily because Richardson-like
convergence can not be defined for discontinuous functions~\cite{Neilsen2005}.
However, it appears here that the constraint violations are propagated
at roughly the same velocity as the out-going shock, and the difference
between free and cleaned evolutions is small.

\begin{table}
\begin{center}
{\bf Spherical blast wave parameters}
\begin{tabular}{c|c|c|c|c|c|c|c|c}
\hline
               &  $\rho_0$ & $P$ & $v^x$ & $v^y$ & $v^z$ & $B^x$ & $B^y$ & $B^z$ \\
  \hline
Inside sphere  &   1.0   & $10^4$ &  0.0  & 0.0   & 0.0   & 4.0   & 0.0   & 0.0  \\
Outside sphere &   1.0   & 0.01   &  0.0  & 0.0   & 0.0   & 4.0   & 0.0   & 0.0  
\end{tabular}
\caption{The initial parameters for the spherical blast wave.  The data 
consists of a uniform fluid background with the pressure set to $10^4$ 
inside a sphere of radius $0.08$ centered at the origin.  The adiabatic index 
is $\Gamma = 4/3$.  The domain of simulation is 
$\{x,y,z\}\in\left[ -1,1\right]$.  The Courant factor is $0.3$.}
\label{table:sphblastwave}
\end{center}
\end{table}

\subsection{Relativistic rotor}
\label{subsec:rr}

The relativistic rotor test case starts with a rigidly rotating fluid and evolves
it in the presence of a magnetic field.  This problem is discussed and examined in $2+1$
dimensions in \cite{Neilsen2005,DelZanna2002rv}.  Here we examine the relativistic rotor
in $3+1$ dimensions, confining the initially rigidly rotating fluid to a sphere
of radius $0.1$ with the angular momentum vector pointing in the $+z$ direction.  The fluid is initially
rotating with an angular velocity of $9.95$.  The initial data and relevant evolution parameters
are given in \tref{table:relrotor}.

\begin{figure}
\begin{center}
\epsfig{file=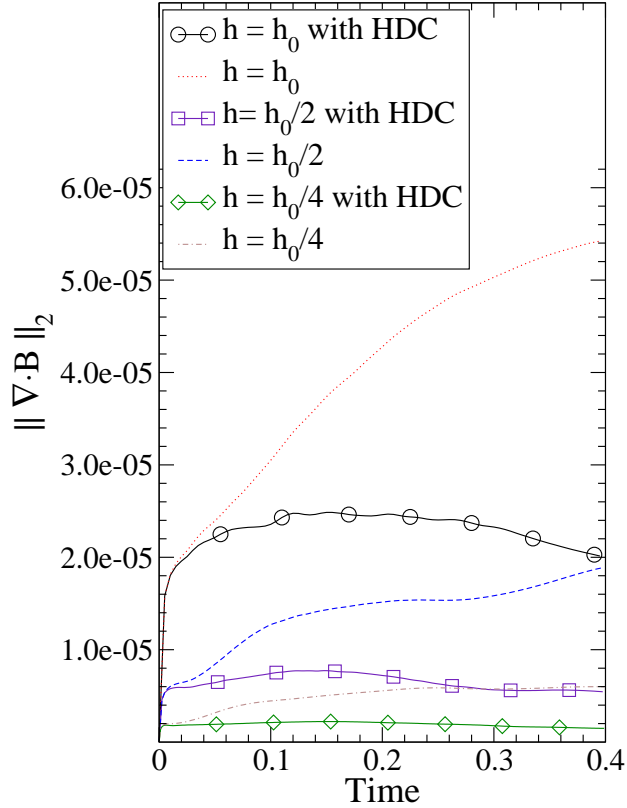,height=12.8cm}
\end{center}
\caption{The $L_2$ norm of the ${\bf \nabla} \cdot {\bf B} = 0$ constraint violation as a function of time for the
relativistic rotor at three resolutions comparing results obtained with free evolution and hyperbolic
divergence cleaning.  Hyperbolic divergence cleaning has a profound impact on constraint
control.  The base resolution is $h_0 = 0.025$.  The hyperbolic divergence cleaning parameters used were $c_h = 1$ and $c_p = 1$.}
\label{fig:relrotor-constraint}
\end{figure}

\begin{figure}
\begin{tabular}{cccc}
&With Divergence Cleaning && Difference\\
  \hline
$P$ & \includegraphics*[height=6.0cm]{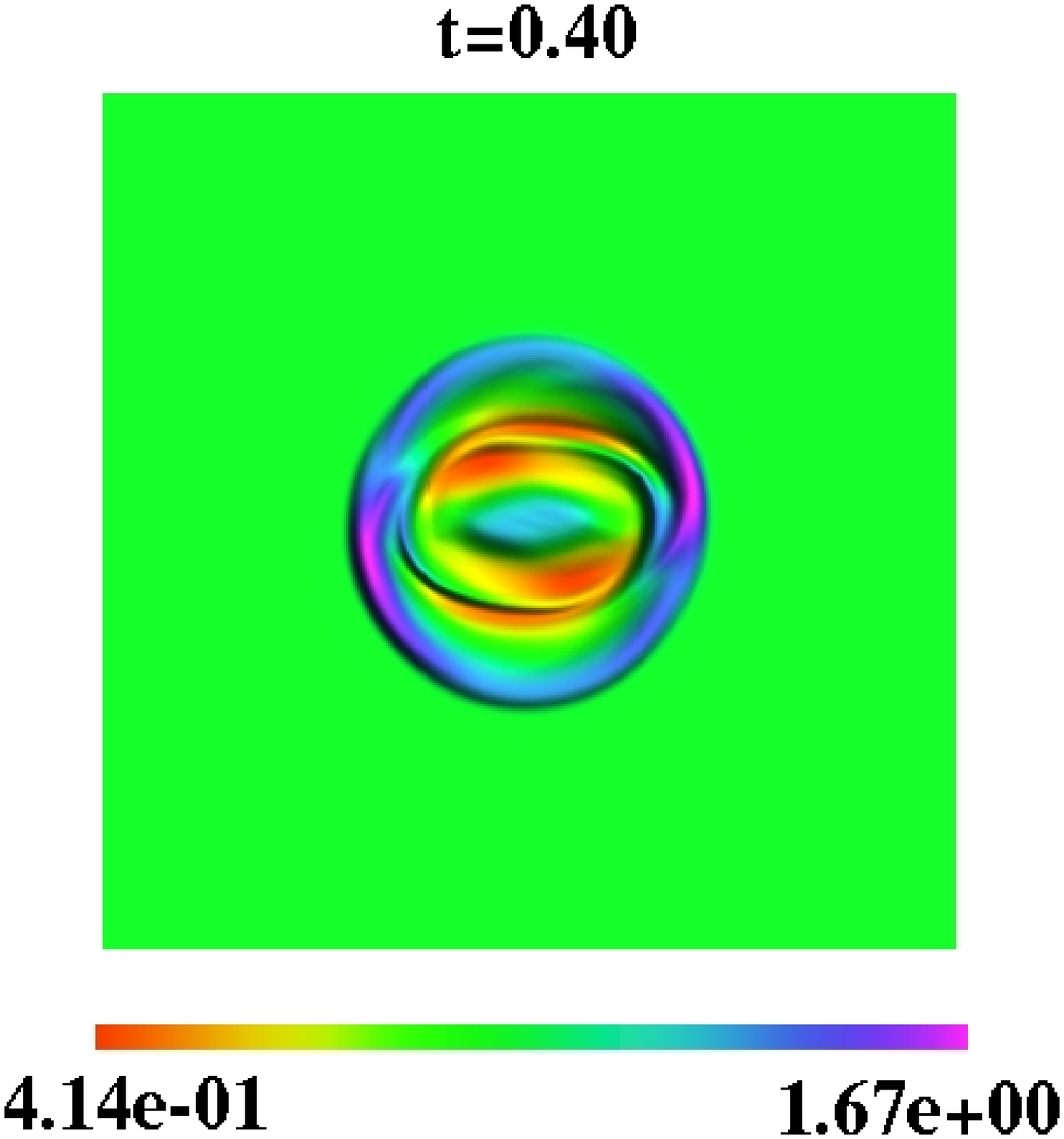}&
$\Delta P$& \includegraphics*[height=6.0cm]{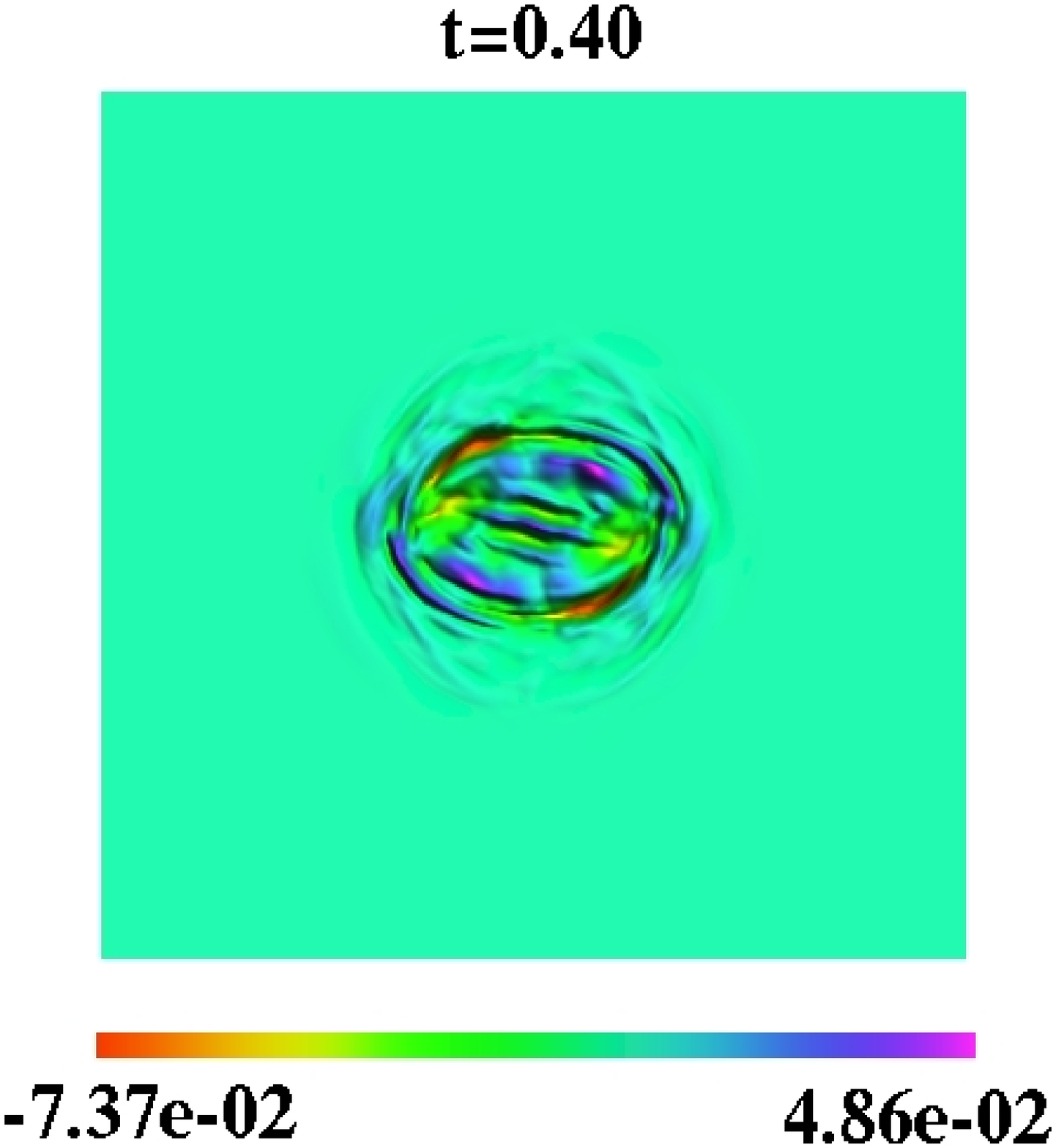}
\end{tabular}
\caption{Unigrid simulation results for the relativistic rotor at time 0.4, showing 
a slice along the $z=0$ plane.  The $x$ axis is the horizontal direction.
The pressure found using hyperbolic divergence cleaning 
is shown on the left.
The difference between the pressure found with and without hyperbolic
divergence cleaning is shown on the right.
This gives an estimation of the relative errors that
arise in free evolutions.  The simulations were performed using a resolution of 
$h=0.00625$ on a domain of $\{x,y,z\}\in\left[ -1,1\right]$.}
\label{fig:2drelrotor}
\end{figure}

Results using a uniform computational grid form the standard against which we measure
the AMR results.  The $L_2$ norms of the  ${\bf \nabla} \cdot {\bf B}$ constraint violation as a function of time 
using three different
unigrid resolutions are shown in \fref{fig:relrotor-constraint}, comparing results obtained both
with and without hyperbolic divergence cleaning.
Hyperbolic divergence cleaning significantly improves the
constraint preservation in the relativistic rotor case.
2-D slices of the pressure
along the $z=0$ plane at time $t=0.4$ are shown in \fref{fig:2drelrotor}.

\begin{figure}
\begin{center}
\epsfig{file=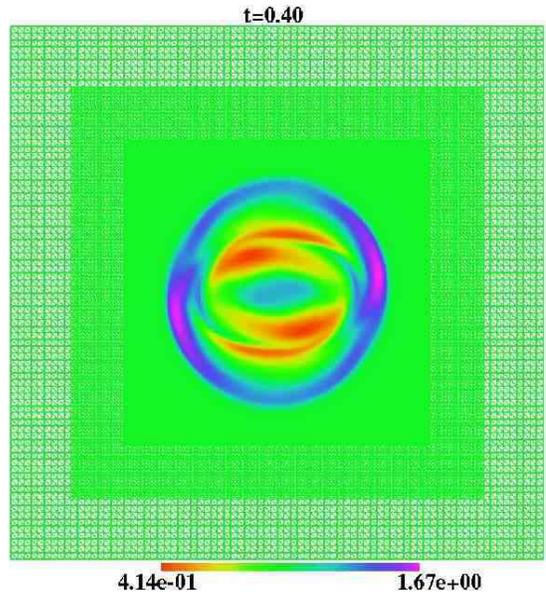,height=8.0cm}
\caption{A wireframe slice along the $z=0$ plane of the
pressure for the relativistic rotor using hyperbolic divergence
 cleaning and two levels of adaptive mesh refinement.  
 The $x$ axis is the horizontal direction.
This simulation reproduces the unigrid solution to better than 0.1\%.
See \fref{fig:ramrdiff}.  This AMR case required five times fewer 
CPU hours
than the comparable unigrid case (shown in \fref{fig:2drelrotor})
and was performed with a maximum resolution of 
$h=0.00625$ on a domain of $\{x,y,z\}\in\left[ -1,1\right]$ 
using 32 processors.}
\label{fig:ramr}
\end{center}
\end{figure}

\begin{figure}
\begin{tabular}{cc}
\epsfig{file=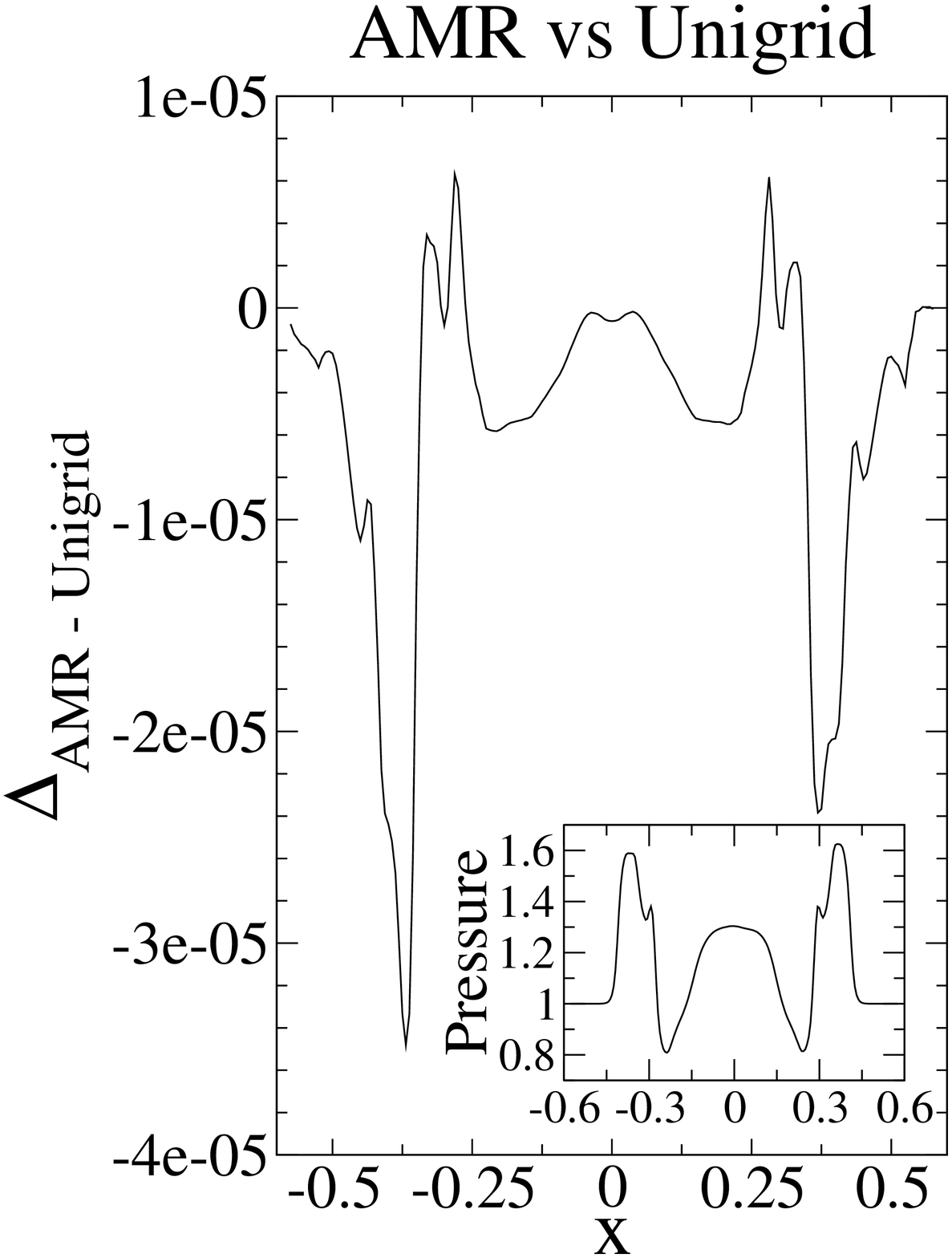,height=8.5cm} & \epsfig{file=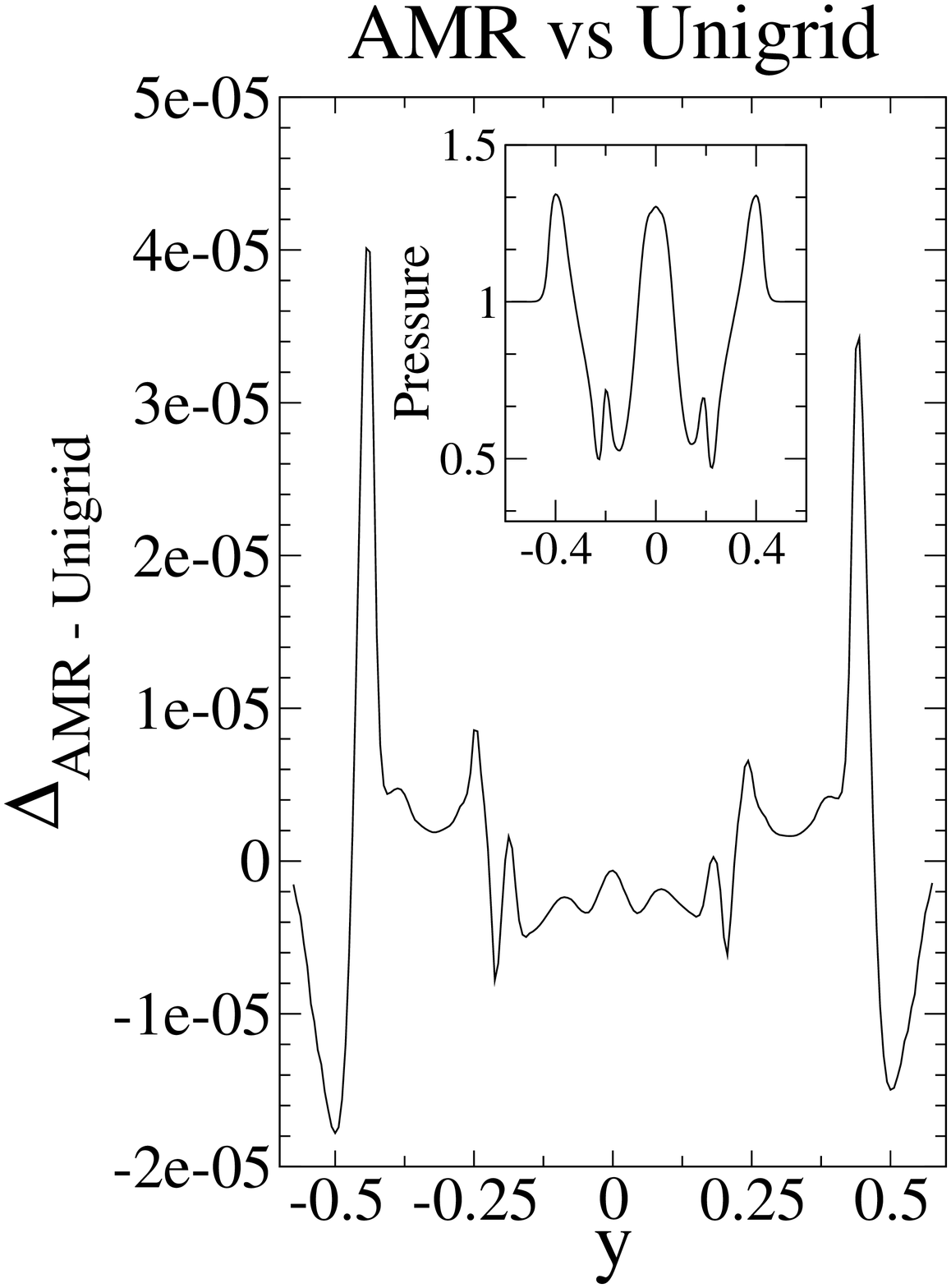,height=8.5cm}  \\
{\bf (a)} & {\bf (b)}
\end{tabular}
\caption{The absolute difference in pressure between
a two level AMR simulation and the equivalent unigrid simulation
for the relativistic rotor case at time $0.4$.
\Fref{fig:ramrdiff}(a) plots the difference in pressure between the AMR and
unigrid results along the $x$ axis; \fref{fig:ramrdiff}(b) plots the difference
along the $y$ axis.  The AMR simulation is the same as that shown
in \fref{fig:ramr}.  The unigrid simulation is the same as that shown
in \fref{fig:2drelrotor}.  The finest resolution meshes of the
two level AMR system had a resolution of $h=0.00625$, equivalent to the
unigrid mesh resolution.  The results are identical to better
than 0.1\%.  For reference, the pressure along the $x$ and $y$ axes are
plotted in the window insets.}
\label{fig:ramrdiff}
\end{figure}

Adaptive mesh refinement results are presented in figures \ref{fig:ramr} and \ref{fig:ramrdiff}.
These figures present a two-level AMR simulation with refinement centered on the shock front.
This AMR simulation required five times fewer CPU hours than the equivalent 
unigrid simulation.  \Fref{fig:ramr} shows the resulting
pressure and mesh hierarchy at time $t=0.4$.  \Fref{fig:ramrdiff} compares the difference
along the $x$ and $y$ axes of the pressure between the AMR and equivalent unigrid simulation.  The
AMR simulation was found to reproduce the unigrid results to well within 0.1\%.

\begin{table}
\begin{center}
{\bf Relativistic rotor parameters}

\begin{tabular}{c|c|c|c|c|c}
\hline
               &  $\rho$ & $P$    & $B^x$ & $B^y$ & $B^z$ \\
  \hline
Inside sphere  &   10.0  & 1.0  & 1.0   & 0.0   & 0.0  \\
Outside sphere &   1.0   & 1.0  & 1.0   & 0.0   & 0.0  
\end{tabular}
\caption{The initial parameters for the relativistic rotor.  The data consists
of an initially rigidly rotating fluid inside a sphere of radius $0.1$ 
centered at the origin with a magnetic field.  The fluid is rotating with $\omega = 9.95$
around the $z$ axis.  The adiabatic index is $\Gamma = 5/3$.  The domain
of simulation is $\left[ -1,1\right]$ in each of the $x$,$y$, and $z$ directions.
The Courant factor used in the presented relativistic rotor simulations was $0.2$.}
\label{table:relrotor}
\end{center}
\end{table}

\subsection{Accretion onto a Black Hole}
\label{subsec:accretion}
Numerical simulation of fluid accretion onto a Schwarzschild black hole
provides a test of the AMR infrastructure using a curved space background
and an excision region.  The steady state solution is given by Michel~\cite{Michel}
and has been explored previously using ingoing Eddington-Finkelstein 
coordinates~\cite{Papadopoulos}.  This solution describes the
continuous spherical accretion of a fluid onto a black hole.

\begin{figure}
\begin{center}
\epsfig{file=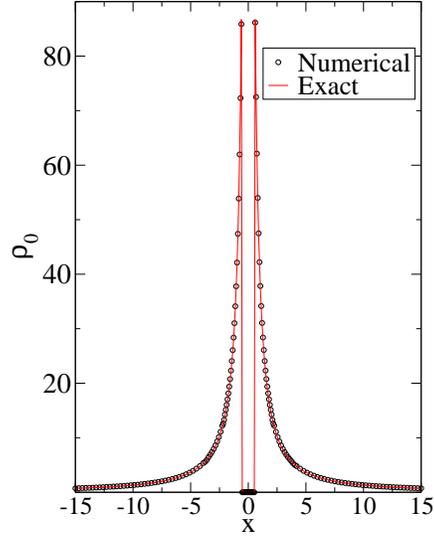,height=8.5cm}
\caption{This figure compares $\rho_0$ from the numerical steady state solution
 of accretion onto a black hole with the Michel solution along the $x-$axis
 of the computational domain.  The fluid
 is initially set to the Michel solution for radius $r > 2.5M$ and
 constant pressure and density for $r \leq 2.5 M$.  The system is
 then evolved until steady state is reached.  The numerical result
 shown here is at $t=50M$ with a finest resolution of $0.075M$.  
The excision region of the black hole is in the center of the grid.  The AMR
grid structure of the simulations is shown in figure \ref{fig:accretion2}.}
\label{fig:accretion}
\end{center}
\end{figure}

\begin{figure}
\begin{tabular}{cc}
\epsfig{file=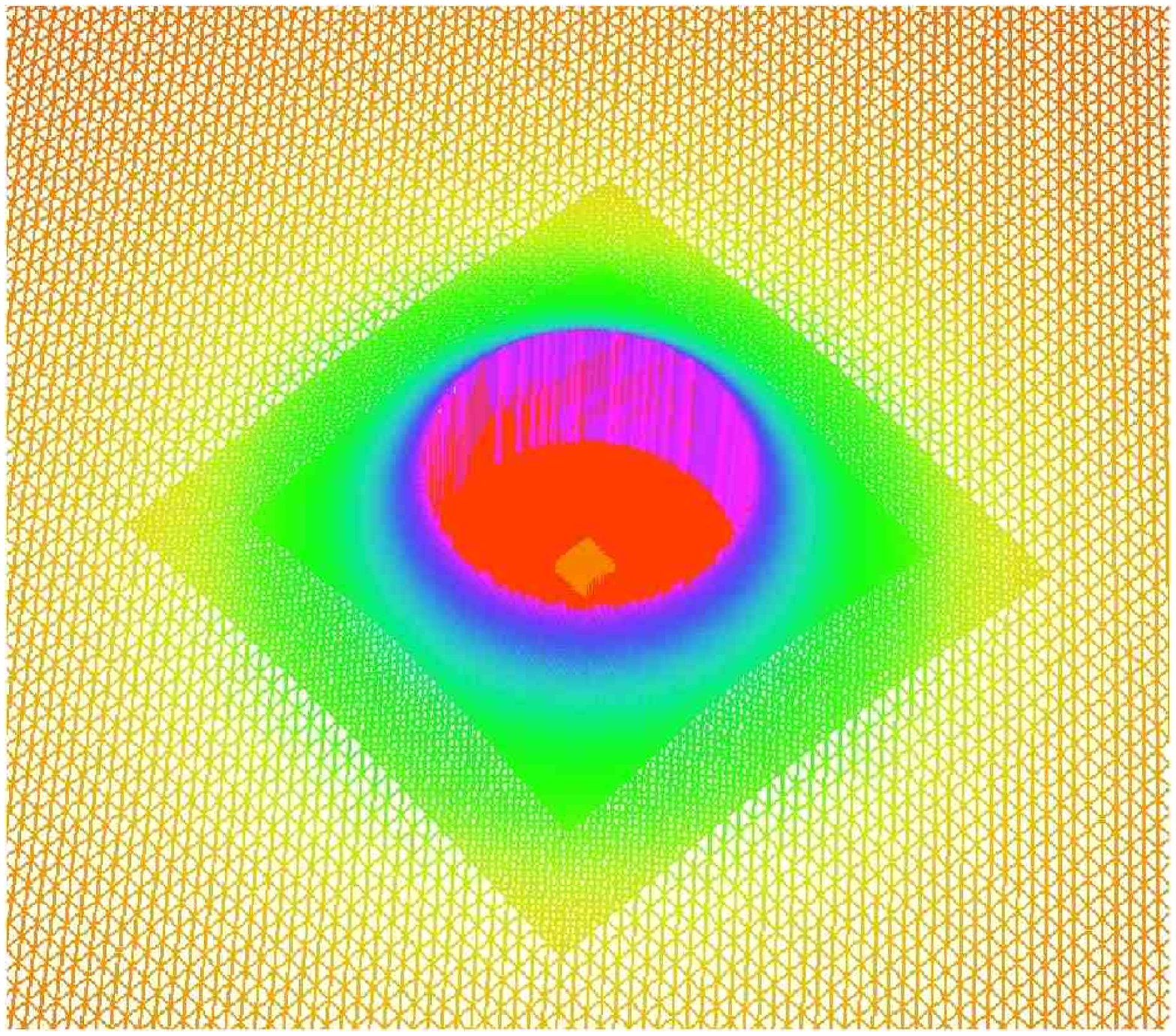,height=5.5cm} & \epsfig{file=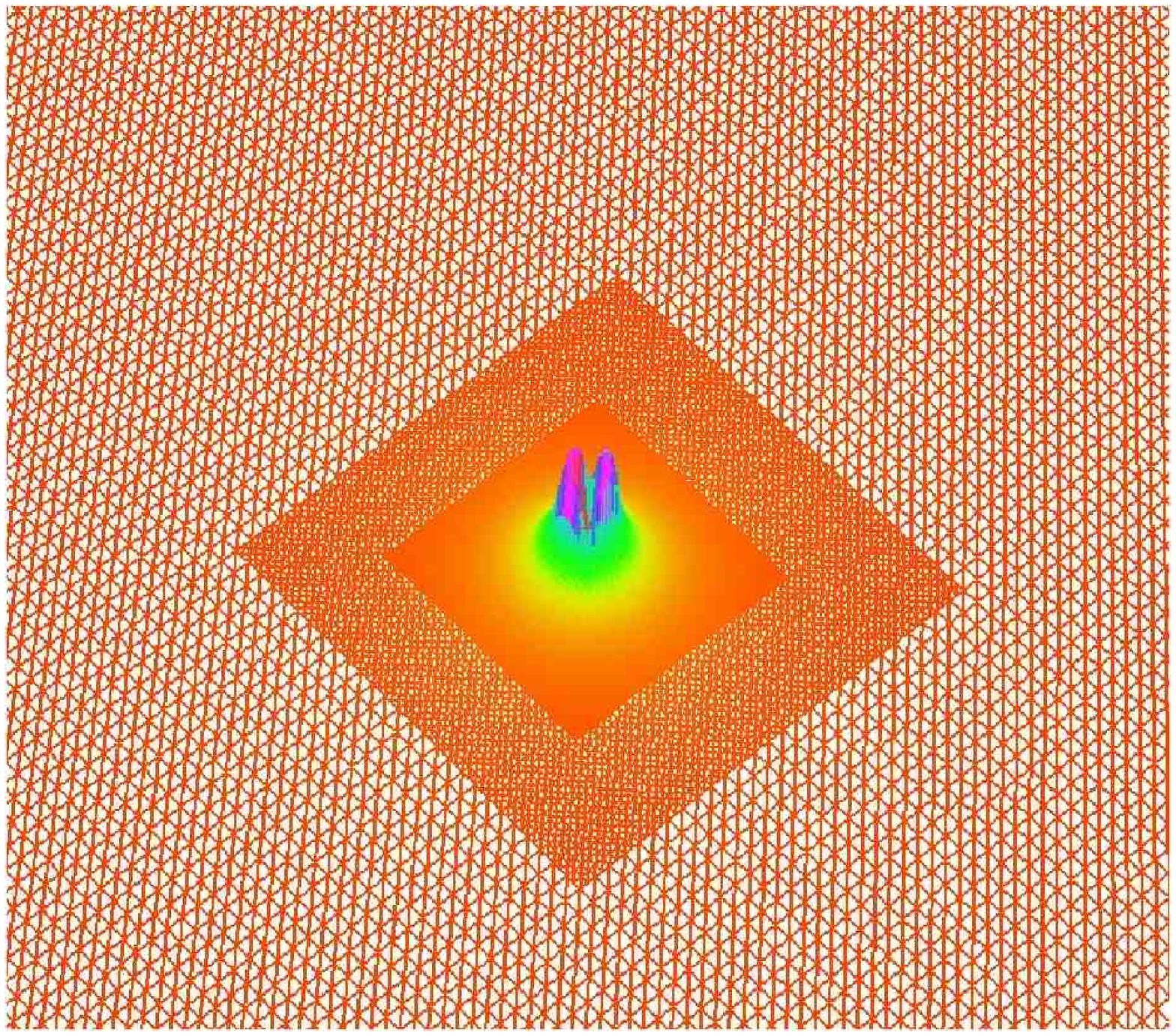,height=5.5cm} \\
{\bf $t = 0M$} & {\bf $t = 50M$} 
\end{tabular}
\caption{This figure shows $\rho_0$ and the AMR grid structures at $t=0M$ and $t=50M$ along the $x-y$ plane.
  The refinement criteria is the shadow hierarchy for truncation error estimation.
 The fluid
 is initially set to the Michel solution for radius $r > 2.5M$ and
 constant pressure and density for $r \leq 2.5 M$.  The system is
 then evolved until steady state is reached.  The cubical excision region
is highlighted in the center of the grid on the left.}
\label{fig:accretion2}
\end{figure}

\begin{figure}
\begin{center}
\epsfig{file=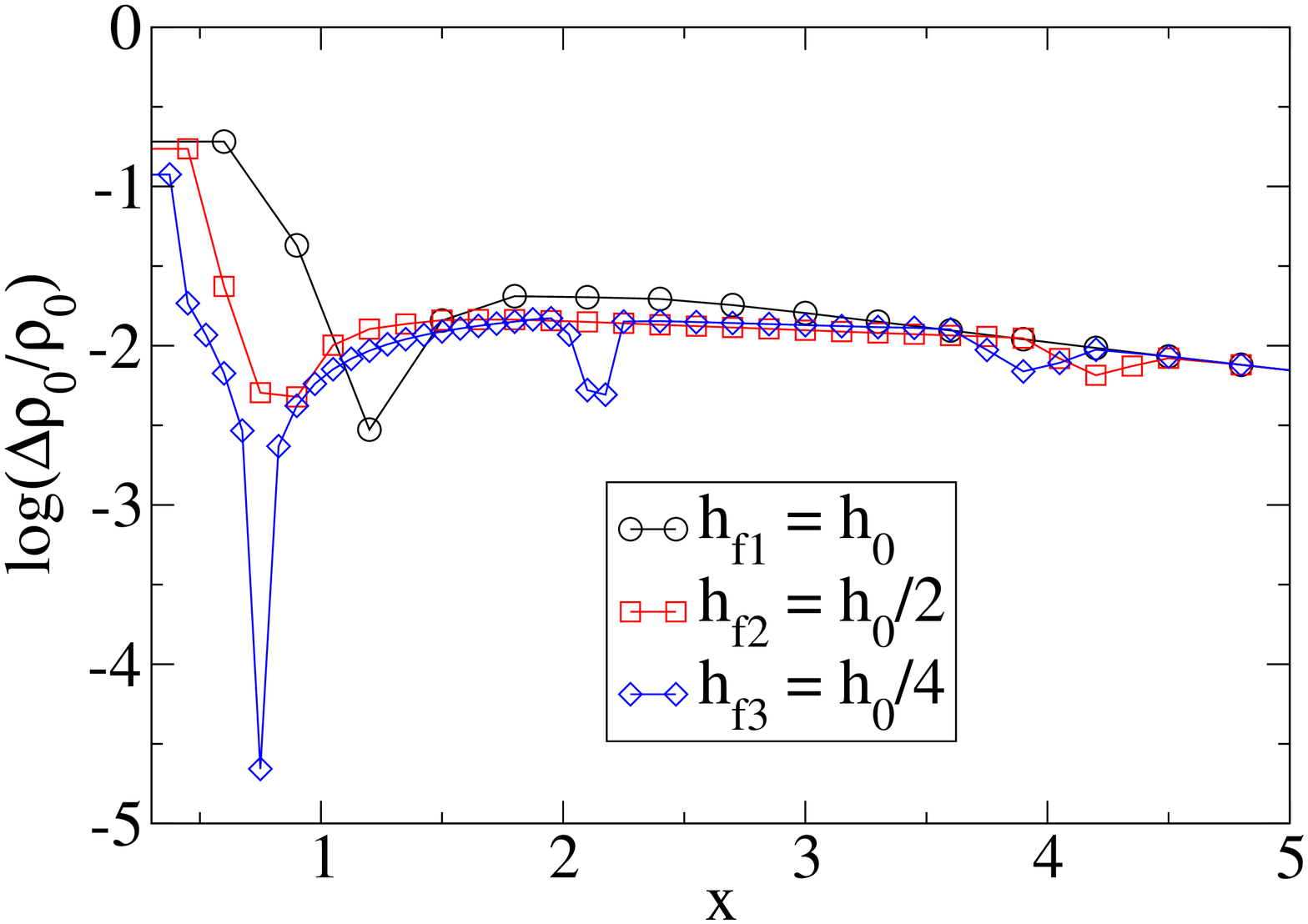,height=8.5cm,angle=270}
\caption{This figure plots $\log(\Delta \rho_0/\rho_0)$ along the $x$-axis 
of the computational domain where $\Delta \rho_0$ is the difference 
in $\rho_0$ between the numerical 
steady state accretion solution and Michel solution at $t=50M$.  
The convergence test consists of successively enabling another 
AMR refinement level to get a finer resolution.  The resolution reported in
the legend of the plot is the finest resolution present in the simulation.  Only for the 
lowest resolution case, $h_{f1}$, is the resolution constant across the entire grid.
Cases $h_{f2}$ and $h_{f3}$ contain multiple resolutions across the domain.
At locations where simulations share the same resolution, 
they also display the same error, modulo AMR boundary effects.  
From $x \in [2.1~M, 15~M]$,
simulations $h_{f2}$ and $h_{f3}$ have the same resolution.  
  From $x \in [3.9~M, 15~M]$ all
  three simulations share the same resolution.  The coarsest resolution for these
 simulations is $h_0 = 0.3~M$.}
\label{fig:accretion3}
\end{center}
\end{figure}

We set a fixed black hole metric using ingoing Eddington-Finkelstein
coordinates.  To avoid the singularity inside the black hole, we implement a cubic excision
region.  
The excision cube is located at the center of the grid and has a half width of $0.3M$.
The boundary condition at the excision region is a copy condition; points next to the
excision region are simply copied into the excision region when necessary for reconstruction.
The mass of the black hole, $M$, is set to one and the black hole is placed at the
center of the grid.  The sonic radius, $r_c$, is selected to be $400M$ with a
density $\rho_c = 0.01$.  The domain of simulation is 
$\{x,y,z\}\in\left[ -15M,15M\right]$.  The Michel steady state solution is found following the
procedure described in~\cite{Papadopoulos} and the outer boundary is kept fixed
at this solution, providing a continual influx of mass onto the black hole.
For radius $r > 2.5M$ the Michel steady state solution is set as
initial data; for $r \leq 2.5M$ the initial data are set to be 
$\rho_0 = 0.1$, $P=0.1$, and $v^i = 0$.  The fluid falls onto the black hole and
eventually reaches steady state.  A comparison of the Michel steady state solution and
numerical solution at $t=50M$ is given in figure \ref{fig:accretion}.
The AMR grid structures at times $t=0M$ and $t=50M$ are given in figure \ref{fig:accretion2}.
A convergence test for $\rho_0$ is presented in figure \ref{fig:accretion3}.


\subsection{Scaling}
\label{subsec:scaling}

\begin{figure}
\begin{tabular}{cc}
\epsfig{file=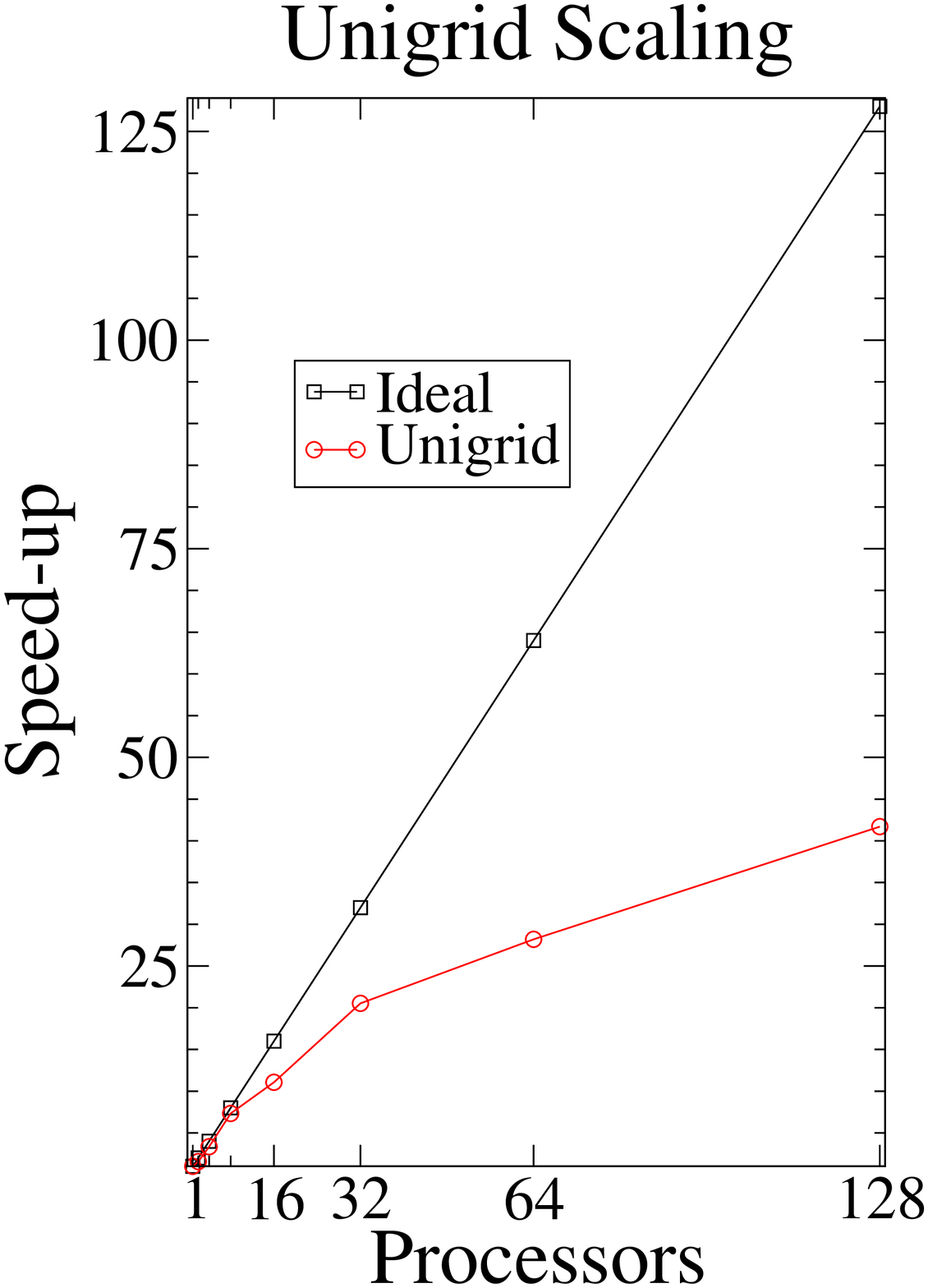,height=8.5cm} & \epsfig{file=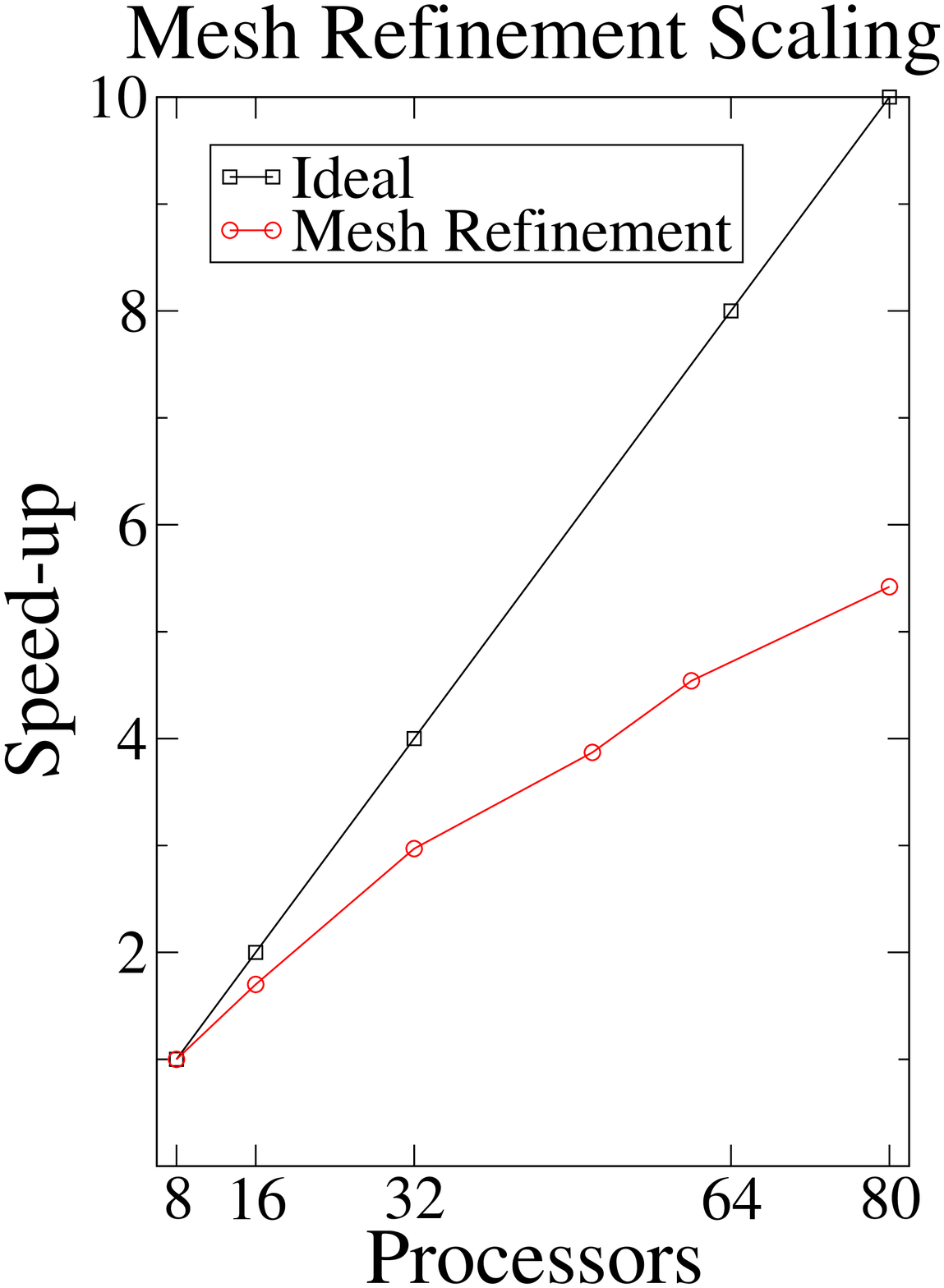,height=8.5cm} \\
{\bf (a)} & {\bf (b)} 
\end{tabular}
\caption{This figure shows strong scaling results for single grid and
 mesh refinement.
The spherical blast wave initial data were run for a fixed problem size
as the number of processors is varied.  The left frame shows 
the unigrid strong scaling results, and the right frame includes mesh 
refinement.  For the unigrid scaling, the data were evolved for 80 iterations,
and the global grid size was $121^3$.  For this problem size, the 
communication overhead begins to overshadow the local process computation 
on $\geq 64$ processors.  For the mesh refinement scaling, 
thirty iterations were performed
on a coarse grid of size $81^3$ and a single level of refinement.  
Since the test problem would not fit in memory on a single processor, 
speed-up was measured using 8 processors as the base value.
All tests were performed on an Intel Pentium IV 3.0 GHz cluster with Myrinet.}
\label{fig:sphshock-strongscaling}
\end{figure}

Unigrid and mesh refinement parallel scaling tests for the spherical blast wave
are given in \fref{fig:sphshock-strongscaling}.  The results presented are the
strong scaling results; the global problem size was kept constant while the 
number of processors varied.  Strong scaling tests are problem dependent and vary  
according to the size of the global problem selected for investigation.  However, 
they also give the most direct indication of performance speed-up
across a wide range of processors for a particular problem.  
Weak scaling tests are where the global problem size is increased
as the number of processors increases in order to keep the problem size local to each processor
constant.  Weak scaling tests were also performed on the spherical blast wave 
problem on 16--256 processors without showing any significant 
performance degradation.  

In the unigrid strong scaling tests of \fref{fig:sphshock-strongscaling}(a) 
a $121^3$ spherical blast
wave problem was evolved for 80 iterations on 1--128 processors.
Speed-up is defined as
    \begin{eqnarray*}
     {\rm speedup}(n) = \frac{{\rm Run~time~on~one~processor}}{{\rm Run~time~on~{\it n}~processors}}.
    \end{eqnarray*}
As the number of processors increases, the communication eventually overshadows the
local processor computation.  For the test problem size examined, this begins on $\geq 64$
processors.  For comparison, strong scaling results 
for a different unigrid TVD MHD code are given in \cite{CITA},
where communication overhead saturation occurs on $> 64$ processors using a $240^3$ mesh size and
$8$ processors the base scaling value.

In the mesh refinement strong scaling tests of \fref{fig:sphshock-strongscaling}(b) 
a $81^3$ spherical blast wave with a single level of mesh refinement was evolved for 30
iterations.  The base number of processors for scaling measurement was $8$ on account of memory
considerations.  Results are shown for 8--80 processors.  

%
%
\section{Conclusion}
\label{sec:conclusion}

We have presented three flat space relativistic MHD tests
and one fluid accretion test using vertex-centered
distributed adaptive mesh refinement and the approximate Riemann solver
algorithm for the GRMHD equations presented in \cite{Neilsen2005}.  
Each of the three relativistic MHD tests, including the Balsara
black wave and the spherical shock and relativistic rotor, have sharp
features requiring high resolutions.
In each of these cases, substantial performance gains of AMR versus
unigrid were observed.  
Two level AMR simulations required between 5--8 times fewer 
CPU hours than the equivalent 
unigrid cases.  AMR results reproduced the unigrid results, often 
better than 0.1\%.

Hyperbolic divergence cleaning was examined in connection with the spherical
blast wave and relativistic rotor cases.     
It had a positive impact 
on constraint control in both cases.
The impact was especially pronounced
in the relativistic rotor case, which has more features interior to the 
outgoing wave front than the spherical blast wave.  
Hyperbolic divergence cleaning worked equally well in both
unigrid and AMR tests.

Fluid accretion onto a Schwarzschild black hole tests the
code using a curved space background and a black hole excision region
as well as using AMR.  In this test, a shadow hierarchy for truncation
error estimation was used as the AMR refinement criterion in 
recovering the known steady state solution.

Parallel
performance measures were presented in connection with the spherical
blast wave.  Speed-ups achieved using the spherical blast wave
were reported both with unigrid and mesh refinement simulations.  Performance
speed-ups were found on up to at least $128$ processors with unigrid and 
up to at least $80$ processors with mesh refinement.  By its very
nature, the strong scaling test is problem dependent: the problem size is
fixed while the number of processors is varied.  This is in contrast to
the weak scaling tests often presented in numerical relativity, where
the problem size per processor is fixed.  Strong scaling tests, however,
address the real-world questions of how long it takes to solve a particular
problem, and how to do it most efficiently.

Having presented these tests, we now turn to some questions of astrophysical
interest mentioned in the introduction.  In particular, we have
included fully dynamical general relativity in our code using the Einstein
equations specified in~\cite{Tiglio:2003xm}.  In 
future work we hope to present evolutions of TOV stars as well as rotating,
magnetized neutron stars.  Recent work suggests interesting effects of 
general relativity with rotation on supermassive polytropes and the bar
mode instability~\cite{Zink:2005rr}.  The addition of magnetic fields to 
these systems may
suggest new questions and provide new insight into magnetic astrophysical 
phenomena. One part of this question includes understanding not only the
interior of a magnetized rotating neutron star, but its magnetosphere as well.
Ideal MHD codes based on Godunov-type schemes frequently encounter 
difficulties when $B^2 \gg \rho_0$, as relatively small truncation errors 
in the evolution of the conserved variables lead to large fractional errors in 
computing the internal energy density and other primitive variables.  While we
have tried to create a robust primitive variable solver, this
difficulty can not be avoided for high-resolution shock-capturing schemes.  
Therefore, a
full  numerical study of such a star and magnetosphere may require coupling 
the equations of ideal MHD for the interior solution with the equations 
of force-free electrodynamics for the exterior.  We are actively pursuing
this question.

Another area to be targeted in future papers is constraint 
preserving boundary conditions for MHD.  Currently we use the conventional 
outflow boundary conditions.  This could be improved by using outer 
boundary conditions 
that are constraint preserving.  Additionally, for some systems we wish to
require that no incoming modes enter the domain.
To this end it would be useful to construct
the full spectral decomposition of our system in order to be able to 
determine ingoing and outgoing modes.  Work in this direction has already
begun and shows promise.

%
%
\ack
We are pleased to thank Luis Lehner, Carlos Palenzuela,
Ignacio Olabarrieta, Patrick Motl, 
Tanvir Rahman, Oscar Reula, and Joel Tohline for helpful discussions 
and comments during the course of this work.   
We also thank Bruno Giacomazzo and Luciano Rezzolla for sharing their
computer code to solve the exact Riemann problem for relativistic MHD.
This work was supported by the National Science Foundation under grants
PHY-0326311 and PHY-0244699 to Louisiana State University, PHY-0326378 and PHY-0502218
to Brigham Young University, and PHY-0325224 to Long Island University.
This research was also supported in part by the National Science Foundation
through TeraGrid resources provided by SDSC under allocation award PHY-040027.
\vspace{10pt}

\bibliography{./mhd}

\begin{thebibliography}{10}

\bibitem{Wilson1975}
J~R Wilson.
\newblock Some magnetic effects in stellar collapse and accretion.
\newblock {\em Ann. New York Acad. Sci.}, 262:123--132, 1975.

\bibitem{Koide:2000}
S~Koide, D~L Meier, K~Shibata, and T~Kudoh.
\newblock General relativistic simulations of early jet formation in a rapidly
  rotating black hole magnetosphere.
\newblock {\em Astrophys. J.}, 536:668, 2000.

\bibitem{DeVilliers:2002ab}
J-P De~Villiers and J~F Hawley.
\newblock A numerical method for general relativistic magnetohydrodynamics.
\newblock {\em Astrophys. J.}, 589:458, 2003.

\bibitem{Gammie:2003rj}
C~F Gammie, J~C McKinney, and G~T\'oth.
\newblock {HARM}: A numerical scheme for general relativistic
  magnetohydrodynamics.
\newblock {\em Astrophys. J.}, 589:444, 2003.

\bibitem{Baumgarte:2002b}
T~W Baumgarte and S~L Shapiro.
\newblock Collapse of a magnetized star to a black hole.
\newblock {\em Astrophys. J.}, 585:930--947, 2003.

\bibitem{Komissarov:2004}
S~S Komissarov.
\newblock General relativistic magnetohydrodynamic simulations of monopole
  magnetospheres of black holes.
\newblock {\em MNRAS}, 350:1431, 2004.

\bibitem{Anton:2005gi}
L~{Ant\'on}, O~Zanotti, J~A Miralles, J~M {Mart\'i}, J~M {Ib\'a\~nez}, J~A
  Font, and J~A Pons.
\newblock Numerical 3+1 general relativistic magnetohydrodynamics: a local
  characteristic approach.
\newblock {\em Astrophys. J.}, 637:296, 2006.

\bibitem{Anninos:2005}
P~Anninos, P~C Fragile, and J~D Salmonson.
\newblock {COSMOS++}: {R}elativistic magnetohydrodynamics on unstructured grids
  with local adaptive refinement.
\newblock {\em Astrophys. J.}, 635:723--740, 2005.

\bibitem{Duez:2005sf}
M~D Duez, Y~T Liu, S~L Shapiro, and B~C Stephens.
\newblock Relativistic magnetohydrodynamics in dynamical spacetimes: Numerical
  methods and tests.
\newblock {\em Phys. Rev. D.}, 72:024029, 2005.

\bibitem{Shibata:2005gp}
M~Shibata and Y~I Sekiguchi.
\newblock Magnetohydrodynamics in full general relativity: Formulation and
  tests.
\newblock {\em Phys. Rev. D.}, 72:044014, 2005.

\bibitem{Balsara2001}
D~S Balsara.
\newblock Total variation diminishing scheme for relativistic
  magnetohydrodynamics.
\newblock {\em ApJS}, 132:83--101, 2001.

\bibitem{Gombosi2000}
T~I Gombosi, D~L De~Zeeuw, C~P~T Groth, and K~G Powell.
\newblock Magnetospheric configuration for {P}arker-spiral {IMF} conditions:
  Results of a 3d {AMR} {MHD} simulation.
\newblock {\em Adv. Space Res.}, 26:139--149, 2000.

\bibitem{Gombosi2001}
T~I Gombosi, G~{T\'oth}, D~L De~Zeeuw, K~G Powell, and Q~F Stout.
\newblock Adaptive mesh refinement {MHD} for global simulations.
\newblock {\em Proceedings of ISSS}, 6:1--8, 2001.

\bibitem{Neilsen2005}
D~Neilsen, E~W Hirschmann, and R~S Millward.
\newblock Relativistic {MHD} and black hole excision: Formulation and initial
  tests.
\newblock {\em Class. Quantum Grav.}, 23:S505--S527, 2006.

\bibitem{Lucas-Serrano:2004aq}
Arturo Lucas-Serrano, Jose~A. Font, Jose~M. Ibanez, and Jose~M. Marti.
\newblock Assessment of a high-resolution central scheme for the solution of
  the relativistic hydrodynamics equations.
\newblock {\em Astron. Astrophys.}, 428:703--715, 2004.

\bibitem{Shibata:2005jv}
Masaru Shibata and Jose~A. Font.
\newblock Robustness of a high-resolution central scheme for hydrodynamic
  simulations in full general relativity.
\newblock {\em Phys. Rev.}, D72:047501, 2005.

\bibitem{motl2006}
P~Motl.
\newblock Introducing {F}low-er: A hydrodynamics code for relativistic and
  {N}ewtonian flows.
\newblock April 2006 APS meeting; slides at
  \verb+http://charybdis.phys.lsu.edu/~patrickmotl/+, 2006.

\bibitem{Dedner2002}
A~Dedner, F~Kemm, D~{Kr\"oner}, C-D Munz, T~Schnitzer, and M~Wesenberg.
\newblock Hyperbolic divergence cleaning for the {MHD} equations.
\newblock {\em J. Comput. Phys.}, 175:645, 2002.

\bibitem{Berger}
M~J Berger and J~Oliger.
\newblock Adaptive mesh refinement for hyperbolic partial differential
  equations.
\newblock {\em J. Comp. Phys.}, 53:484, 1984.

\bibitem{Lehner2006}
L~Lehner, S~L Liebling, and O~Reula.
\newblock {AMR}, stability and higher accuracy.
\newblock {\em Class. Quant. Grav.}, 23:S421--S446, 2006.

\bibitem{ShuOsherI}
C-W Shu and S~Osher.
\newblock Efficient implementation of essentially non-oscillatory
  shock-capturing schemes.
\newblock {\em J. Comput. Phys.}, 77(2):439--471, 1988.

\bibitem{Baumgarte:2002a}
T~W Baumgarte and S~L Shapiro.
\newblock General relativistic mhd for the numerical construction of dynamical
  spacetimes.
\newblock {\em Astrophys. J.}, 585:921--929, 2003.

\bibitem{Sloan:1985}
J~Sloan and L~L Smarr.
\newblock {\em Numerical Astrophysics}, page~52.
\newblock Jones and Bartlett, 1985.

\bibitem{Evans:1988}
C~R Evans and J~F Hawley.
\newblock Simulation of magnetohydrodynamic flows -- a constrained transport
  method.
\newblock {\em Astrophys. J.}, 332:659, 1988.

\bibitem{Zhang:1989}
X~H Zhang.
\newblock 3+1 formulation of general-relativistic perfect magnetohydrodynamics.
\newblock {\em Phys. Rev. D}, 39:2933, 1989.

\bibitem{Anile}
A~M Anile.
\newblock {\em Relativistic fluids and magneto-fluids}.
\newblock Cambridge University Press, 1989.

\bibitem{Komissarov:1999}
S~S Komissarov.
\newblock Godunov-type scheme for relativistic magnetohydrodynamic.
\newblock {\em MNRAS}, 303:343, 1999.

\bibitem{DelZanna2002rv}
L~Del~Zanna, N~Bucciantini, and P~Londrillo.
\newblock An efficient shock-capturing central-type scheme for multidimensional
  relativistic flows. {II}. {M}agnetohydrodynamics.
\newblock {\em Astron. Astrophys.}, 400:397--414, 2003.

\bibitem{Leismann:2005}
T~Leismann, L~{Ant\'on}, M~A Aloy, {M\"uller} E, J~M {Mart\'i}, J~A Miralles,
  and J~M {Ib\'a\~nez}.
\newblock Relativistic mhd simulations of extragalactic jets.
\newblock {\em Astron. and Astrophys.}, 436:503--526, 2005.

\bibitem{LiuOsher}
X-D Liu and S~Osher.
\newblock Convex {ENO} high order multi-dimensional schemes without field by
  field decomposition or staggered grids.
\newblock {\em J. Comp. Phys.}, 142:304--330, 1998.

\bibitem{DelZanna:2002qr}
L~Del~Zanna and N~Bucciantini.
\newblock An efficient shock-capturing central-type scheme for multidimensional
  relativistic flows. {I}. {H}ydrodynamics.
\newblock {\em A\&A}, 390:1177--1186, 2002.

\bibitem{Harten}
A~Harten, P~D Lax, and B~van Leer.
\newblock On upstream differencing and {G}odunov-type schemes for hyperbolic
  conservation laws.
\newblock {\em SIAM Rev.}, 25:35--61, 1983.

\bibitem{BrackbillBarnes}
J~U Brackbill and D~C Barnes.
\newblock The effect of nonzero $\nabla \cdot${{\bf B}} on the numerical
  solution of the magnetohydrodynamic equations.
\newblock {\em J. Comput. Phys.}, 35:426, 1980.

\bibitem{Brackbill}
J~U Brackbill.
\newblock Fluid modeling of magnetized plasmas.
\newblock {\em Space Sci. Rev.}, 42:153--167, 1985.

\bibitem{Balsara2001b}
D~S Balsara.
\newblock Divergence-free adaptive mesh refinement for magnetohydrodynamics.
\newblock {\em J. Comput. Phys.}, 174:614, 2001.

\bibitem{TothRoe}
G~{T\'oth} and P~L Roe.
\newblock Divergence- and curl-preserving prolongation and restriction
  formulas.
\newblock {\em J. Comput. Phys.}, 180:736--750, 2002.

\bibitem{LiLi}
S~Li and H~Li.
\newblock A novel approach of divergence-free reconstruction for adaptive mesh
  refinement.
\newblock {\em J. Comput. Phys.}, 161:605--652, 2000.

\bibitem{Balsara2004}
D~S {Balsara}.
\newblock {Second-order-accurate schemes for magnetohydrodynamics with
  divergence-free reconstruction}.
\newblock {\em ApJS}, 151:149--184, March 2004.

\bibitem{Toth}
G~{T\'oth}.
\newblock The $\nabla\cdot$ {{\bf B}} $=0$ constraint in shock-capturing
  magnetohydrodynamics codes.
\newblock {\em J. Comput. Phys.}, 161:605--652, 2000.

\bibitem{BalsaraKim}
D~S Balsara and J~Kim.
\newblock A comparison between divergence-cleaning and staggered-mesh
  formulations for numerical magnetohydrodynamics.
\newblock {\em Ap. J.}, 602:1079--1090, 2004.

\bibitem{Brodbeck1999}
O~Brodbeck, S~Frittelli, P~Hubner, and O~A Reula.
\newblock Einstein's equations with asymptotically stable constraint
  propagation.
\newblock {\em J. Math. Phys.}, 40:909, 1999.

\bibitem{Palenzuela}
E~Hirschmann, L~Lehner, D~Neilsen, C~Palenzuela, and O~Reula.
\newblock Paper in preparation.

\bibitem{SebastianShu}
K~Sebastian and C-W Shu.
\newblock Multidomain {WENO} finite difference method with interpolation at
  subdomain interfaces.
\newblock {\em J. Sci. Comput.}, 19:405, 2003.

\bibitem{Liebling}
S~L Liebling.
\newblock The singularity threshold of the nonlinear sigma model using 3d
  adaptive mesh refinement.
\newblock {\em Phys. Rev. D}, 66:041703, 2002.

\bibitem{SAMRAI}
R~Hornung, S~Kohn, N~Elliott, S~Smith, A~Wissink, B~Gunney, and D~Hysom.
\newblock {SAMRAI} home page.
\newblock http://www.llnl.gov/CASC/SAMRAI/, 2006.

\bibitem{Chombo}
Lawrence Berkeley National~Lab Applied Numerical Algorithms~Group.
\newblock Chombo --infrastructure for adaptive mesh refinement.
\newblock http://seesar.lbl.gov/anag/chombo/, 2006.

\bibitem{Paramesh}
P~MacNeice, K~Olson, C~Mobarry, R~deFainchtein, and C~Packer.
\newblock {PARAMESH} : A parallel adaptive mesh refinement community toolkit.
\newblock {\em Computer Physics Communications}, 126:33--354, 2000.

\bibitem{AMROC}
R~Deiterding.
\newblock {AMROC} home page.
\newblock http://amroc.sourceforge.net/, 2006.

\bibitem{Boxlib}
M~Lijewki, V~Beckner, and C~Rendleman.
\newblock Boxlib home page.
\newblock http://seesar.lbl.gov/ccse/Software/index.html, 2006.

\bibitem{Rigoutsos}
M~Berger and I~Rigoutsos.
\newblock An algorithm for point clustering and grid generation.
\newblock {\em IEEE Trans. on Systems, Man, and Cybernetics}, 21:1278--1286,
  1991.

\bibitem{Rendleman}
C~Rendleman, V~Beckner, M~Lijewski, W~Crutchfield, and J~Bell.
\newblock Parallelization of structured, hierarchical adaptive mesh refinement
  algorithms.
\newblock {\em Computing and Visualization in Science}, 3:147--157, 2000.

\bibitem{Pretorius}
F~Pretorius.
\newblock {\em Numerical simulations of gravitational collapse}.
\newblock PhD thesis, The University of British Columbia, 2002.

\bibitem{Giacomazzo2005jy}
B~Giacomazzo and L~Rezzolla.
\newblock The exact solution of the {R}iemann problem in relativistic {MHD}.
\newblock {\em J. Fluid Mech.}, 562:223--259, 2006.

\bibitem{Michel}
F~Michel.
\newblock Accretion of matter by condensed objects.
\newblock {\em Astrophys. Space Sci.}, 15:153, 1972.

\bibitem{Papadopoulos}
P~Papadopoulos and J~Font.
\newblock Relativistic hydrodynamics around black holes and horizon adapted
  coordinate systems.
\newblock {\em Phys. Rev. D}, 61:024015, 2000.

\bibitem{CITA}
H~Merz.
\newblock Scaling performance of {CubePM} and {TVD} {MHD} on blue gene.
\newblock
  http://www-03.ibm.com/servers/deepcomputing/bluegene\_literature.html, 2006.

\bibitem{Tiglio:2003xm}
M~Tiglio, L~Lehner, and D~Neilsen.
\newblock 3d simulations of {E}instein's equations: symmetric hyperbolicity,
  live gauges and dynamic control of the constraints.
\newblock {\em Phys. Rev.}, D70:104018, 2004.

\bibitem{Zink:2005rr}
B~Zink et~al.
\newblock Black hole formation through fragmentation of toroidal polytropes.
\newblock {\em Phys. Rev. Lett.}, 96:161101, 2006.

\end{thebibliography}
\bibliographystyle{unsrt}

\end{document}